\documentclass[aps,10pt,pre,twocolumn]{revtex4-1}

\usepackage{amsmath,amssymb,amsfonts,amsthm}
\usepackage[utf8]{inputenc}
\usepackage{graphicx}
\usepackage[caption=false]{subfig}
\usepackage{bbm}
\usepackage[pdftex,bookmarks=false,colorlinks=true,linkcolor=blue,
citecolor=blue,filecolor=black,urlcolor=blue]{hyperref}

\theoremstyle{plain} 
\newtheorem{definition}{Definition}

\newtheorem{lemma}[definition]{Lemma}

\providecommand{\ud}{\mathrm{d}}
\providecommand{\abs}[1]{\left\lvert#1\right\rvert}
\providecommand{\bra}[1]{\left\langle#1\right\vert}
\providecommand{\ket}[1]{\left\vert#1\right\rangle}

\begin{document}

\title{Dynamics of the Bose-Hubbard Chain for Weak Interactions}

\author{Martin L.R. F\"urst}
\email{mfuerst@ma.tum.de}
\affiliation{Excellence Cluster Universe,
Technische Universit\"at M\"unchen,
Boltzmannstra{\ss}e 2}
\affiliation{Mathematics Department,
Technische Universit\"at M\"unchen,
Boltzmannstra{\ss}e 3,
85748 Garching bei M\"unchen,
Germany}

\author{Christian B. Mendl}
\email{mendl@ma.tum.de}
\affiliation{Mathematics Department,
Technische Universit\"at M\"unchen,
Boltzmannstra{\ss}e 3}

\author{Herbert Spohn}
\email{spohn@ma.tum.de}
\affiliation{Mathematics Department,
Technische Universit\"at M\"unchen,
Boltzmannstra{\ss}e 3}
\affiliation{Physics Department,
Technische Universit\"at M\"unchen,
James-Franck-Stra{\ss}e 1}

\date{\today}

\begin{abstract}
\noindent We study the Boltzmann transport equation for the Bose-Hubbard chain in the kinetic regime. The time-dependent Wigner function is matrix-valued with odd dimension due to integer spin. For nearest neighbor hopping only, there are infinitely many additional conservation laws and nonthermal stationary states. Adding longer range hopping amplitudes entails exclusively thermal equilibrium states. We provide a derivation of the Boltzmann equation based on the Hubbard hamiltonian, including general interactions beyond on-site, and illustrate the results by numerical simulations. In particular, convergence to thermal equilibrium states with negative temperature is investigated.
\end{abstract}

\maketitle

\section{Introduction}

In recent years, Bose-Hubbard models have been realized in experiments using ultracold bosonic atoms in optical lattices \cite{BlochDalibardZwergerReview2008, BlochDalibardNascimbene2012}. These experiments facilitate the study of many-body effects like phase transitions from a superfluid to a Mott insulator \cite{SFMott2002} and the (de-) coherence dynamics induced by the Hubbard model \cite{DecoherenceDynamics2007, CoherenceDynamics2007, BlochReview2008}. Nevertheless, the out-of-equilibrium dynamics, convergence to equilibrium and the dynamics after a sudden quench remain topics of active research \cite{NewtonCradle2006, QuenchDynamics2007, NonequilibriumDynamicsBoseHubbard2011}.

In this contribution, we study the dynamics of the Bose-Hubbard chain in the \emph{weakly interacting} (``superfluid'') regime, described by kinetic theory. Our formalism allows for general hopping amplitudes (nearest neighbor, next-nearest neighbor etc.) and interactions beyond solely on-site interactions. We use \cite{BoltzmannFermi2012, BoltzmannNonintegrable2013, DerivationBoltzmann2013} on Fermi-Hubbard as a blueprint. But the details, both theoretical and numerical, differ. In view of the importance of the Bose-Hubbard model a separate study will be of use.

We establish that, for nearest neighbor hopping, on the kinetic level there are infinitely many conservation laws (in addition to the standard density and energy conservation), and consequently nonthermal stationary states. We characterize these stationary states and establish a one-to-one mapping to the conserved quantities.

The additional conservation laws disappear when turning on couplings beyond nearest neighbor hopping: all stationary states are thermal (Bose-Einstein) distributions. For small next-nearest neighbor hopping amplitudes, we observe a prethermalization effect \cite{PrethermalizationEckstein2011, PrethermalizationSchmiedmayer2012} with two time scales, where the system converges quickly to a quasistationary nonthermal state and then relaxes slowly to thermal equilibrium.

Our formalism allows for negative temperatures, as recently realized experimentally \cite{NegativeAbsoluteTemperature2013}. We will illustrate by a model calculation in Sec.~\ref{sec:Simulation} that shifting the momentum of the initial Wigner state, $k \to k + \frac{1}{2}$, flips the sign of the temperature of the ($t \to \infty$) stationary thermal state. Interestingly, this thermal state is (in general) not simply a shifted copy of the thermal state matching the initial state before the shift.

While outside the scope of our contribution, we have to point out one important feature of the kinetic equation for the Bose-Hubbard model. Physically, for dimension $d \ge 3$ and at sufficiently high density there will be a superfluid phase, a property which is still reflected at the kinetic level, see \cite{SpohnBEC2010} and references therein. In the spatially homogeneous setting, if the initial Wigner function is smooth but of a sufficiently high density, after some finite time-span a $\delta$-function will be formed at momentum $k = 0$. The kinetic equation has then to be augmented by coupling it to an evolution equation for the superfluid density. For $d = 1$, as discussed here, to each initial Wigner function there is a uniquely determined stationary Bose-Einstein distribution. For $d \ge 3$, this property holds only if the superfluid density is included.

\section{Bose-Hubbard hamiltonian}
\label{sec:Hamiltonian}

We first write down the hamiltonian of the Bose-Hubbard chain under study. The bosons are described by an integer spin-$n$ field on $\mathbb{Z}$ with creation and annihilation operators satisfying the commutation relations
\begin{align}
&[ a_\sigma(x)^*,  a_\tau(y)\   ] = \delta_{xy} \delta_{\sigma \tau},\\
&[ a_\sigma(x),\   a_\tau(y)\ \:] = 0,\\
&[ a_\sigma(x)^*,  a_\tau(y)^*  ] = 0
\end{align}
for $x, y \in \mathbb{Z}$, $\sigma, \tau \in \{-n, \dots, n\}$, and $[ A, B ] = A B - B A$. The hamiltonian reads
\begin{equation}
\begin{split}
\label{eq:BoseHubbardHamiltonian}
H &= H_0 + \lambda H_1 \\
&= \sum_{x, y \in \mathbb{Z}} \alpha(x - y) \, a(x)^* \cdot a(y) \\
&\quad + \frac{\lambda}{2} \sum_{x, y \in \mathbb{Z}} V(x-y) \big(a(x)^* \cdot a(x)\big) \big(a(y)^* \cdot a(y)\big).
\end{split}
\end{equation}
Here $\alpha$ is the hopping amplitude, which satisfies $\alpha(x) = \alpha(x)^*$ and $\alpha(x) = \alpha(-x)$. The dispersion relation $\omega(k)$ is precisely its Fourier transform: $\omega(k) = \hat{\alpha}(k)$. In Eq.~\eqref{eq:BoseHubbardHamiltonian}, $a(x)^* \cdot a(x) = \sum_{\sigma} a_{\sigma}(x)^* \, a_{\sigma}(x)$, and $0 < \lambda \ll 1$ is the strength of the interaction. The pair potential $\lambda V$ consists of a scalar-valued nonnegative function $V: \mathbb{Z} \to \mathbb{R}$ which satisfies $V(x) = V(-x)$. For the on-site case, $V(x) = \delta_{x,0}$, the Fourier transform is constant, $\hat{V}(k) \equiv 1$.

We use the following convention for the Fourier transform:
\begin{equation}
\label{eq:Fouriertransformation}
\hat{f}(k) = \sum_{x \in \mathbb{Z}} f(x)\, \mathrm{e}^{-2 \pi \mathrm{i}\, k \, x},
\end{equation}
such that the first Brillouin zone is the interval $\mathbb{T} = [-\tfrac{1}{2}, \tfrac{1}{2}]$ with periodic boundary conditions. $H$ can be written in Fourier space as
\begin{equation}
\begin{split}
H = &\int_{\mathbb{T}} \ud k \, \omega(k) \, \big( \hat{a}(k)^* \cdot \hat{a}(k) \big) \\
&+ \frac{\lambda}{2} \int_{\mathbb{T}^4} \ud^4 \boldsymbol{k} \, \delta(\underline{k}) \, \hat{V}(k_1-k_2) \\
&\qquad\quad \times \big(\hat{a}(k_1)^* \cdot \hat{a}(k_2) \big)\, \big(\hat{a}(k_3)^* \cdot \hat{a}(k_4)\big)
\end{split}
\end{equation}
with $\underline{k} = k_1 - k_2 + k_3 - k_4 \mod 1$ and $\ud^4 \boldsymbol{k} = \ud k_1\, \ud k_2\, \ud k_3\, \ud k_4$. Note that the convention for $\underline{k}$ differs from \cite{BoltzmannFermi2012, BoltzmannNonintegrable2013} by an interchange of $k_2 \leftrightarrow k_3$, for consistency with the derivation in Sec.~\ref{sec:Derivation}.

In this contribution we will study a prototypical model with nearest neighbor hopping and an additional next-nearest neighbor hopping term with tunable weight $\eta$. The corresponding dispersion relation reads
\begin{equation}
\label{eq:omega_nnn}
\omega_{\eta}(k) = 1 - \cos(2 \pi k) - \eta \cos(4 \pi k),
\end{equation}
and the pure nearest neighbor hopping case corresponds to $\eta = 0$.

\section{Boltzmann-Hubbard equation}
\label{sec:BoltzmannHubbard}

We will derive the kinetic Boltzmann equation in section~\ref{sec:Derivation}, in analogy to the fermionic case \cite{DerivationBoltzmann2013}. The central object is the two-point function $W(k,t)$ defined by the relation
\begin{equation}
\langle \hat{a}_\sigma(k,t)^* \, \hat{a}_\tau(\tilde{k},t) \rangle = \delta(k - \tilde{k})\, W(k,t)_{\sigma \tau}.
\end{equation}
For each $k \in \mathbb{T}$, $W(k,t)$ is a $(2n+1) \times (2n+1)$ positive semidefinite matrix. The resulting Boltzmann equation reads
\begin{equation}
\label{eq:BoltzmannEquation}
\frac{\partial}{\partial t} W(k,t)
= \mathcal{C}_\mathrm{c}[W](k,t) + \mathcal{C}_\mathrm{d}[W](k,t) = \mathcal{C}[W](k,t),
\end{equation}
with the first term of Vlasov type,
\begin{equation}
\label{eq:Cc}
\mathcal{C}_\mathrm{c}[W](k,t) = - \mathrm{i}\,[H_\mathrm{eff}(k,t), W(k,t)],
\end{equation}
where the effective hamiltonian $H_\mathrm{eff}(k,t)$ is a $(2n+1) \times (2n+1)$ matrix which itself depends on $W$. More explicitly,
\begin{multline}
\label{eq:Heff}
H_{\mathrm{eff},1} = \int_{\mathbb{T}^3} \ud k_2 \ud k_3 \ud k_4 \, \delta(\underline{k}) \, \mathcal{P}\!\left(\tfrac{1}{\underline{\omega}}\right)\\
\times \Big( \hat{V}_{2 3} \hat{V}_{3 4} \left( W_2 W_3 + W_3 W_2 - W_2 W_4 + W_3 \right) \\
+ \hat{V}_{3 4}^2 \, \mathrm{tr}\!\left[W_3 - W_4\right] W_2 \Big).
\end{multline}
Here and later on we use the shorthand $\tilde{W} = 1 + W$, $W_1 = W(k_1,t)$, $H_{\mathrm{eff},1} = H_\mathrm{eff}(k_1,t)$, $\underline{\omega} = \omega(k_1) - \omega(k_2) + \omega(k_3) - \omega(k_4)$, and $\hat{V}_{ij} = \hat{V}(k_i - k_j)$. Note that $\hat{V}_{3 4} = \hat{V}_{1 2}$ in Eq.~\eqref{eq:Heff} due to $k_1 - k_2 = k_4 - k_3$ and the symmetry of $\hat{V}$.

The collision term $\mathcal{C}_\mathrm{d}$ can be written as
\begin{multline}
\label{eq:Cd}
\mathcal{C}_\mathrm{d}[W]_1 = \pi \int_{\mathbb{T}^3} \ud k_2 \ud k_3 \ud k_4 \, \delta(\underline{k}) \, \delta(\underline{\omega}) \\
\times \big( \mathcal{A}[W]_{1234} + \mathcal{A}[W]_{1234}^* \big),
\end{multline}
where the index $1234$ means that the matrix $\mathcal{A}[W]$ depends on $k_1$, $k_2$, $k_3$, and $k_4$. Explicitly
\begin{multline}
\label{eq:AW1234def}
\mathcal{A}[W]_{1234} = \hat{V}_{23} \hat{V}_{34} \, W_4 \tilde{W}_3 W_2 + \hat{V}_{34}^2 \, W_4 \, \mathrm{tr}[W_2 \tilde{W}_3] \\
+ W_1 \Big( \hat{V}_{23} \hat{V}_{34} \big( W_2 \tilde{W}_4 - W_3 \tilde{W}_4 - W_2 \tilde{W}_3 \big) \\
+ \hat{V}_{34}^2 \big( \tilde{W}_4 \, \mathrm{tr}[W_2 - W_3] - \mathrm{tr}[W_2 \tilde{W}_3] \big) \Big).
\end{multline}
The ``gain term'' consisting of the first two summands (plus their conjugate-transposes) is always positive semidefinite, such that the collision operator pushes an hypothetical zero eigenvalue of $W_1$ back to positive values. (The term $W_1 (\dots)$ projected onto the corresponding eigenvector vanishes in this case.) The positivity of the gain term is discussed in appendix~\ref{sec:Positivity}.

Using $k_2 \leftrightarrow k_4$, the integrand in Eq.~\eqref{eq:Cd} admits the reformulation
\begin{equation}
\mathcal{A}[W]_{1234} + \mathcal{A}[W]_{1234}^* = \mathcal{A}_{\mathrm{quad}}[W]_{1234} + \mathcal{A}_{\mathrm{tr}}[W]_{1234}
\end{equation}
with
\begin{equation}
\label{eq:Aquad}
\begin{split}
\mathcal{A}_{\mathrm{quad}}[W]_{1234} &\\
= \hat{V}_{23} \hat{V}_{34} \Big(
& + \tilde{W}_1 W_2 \tilde{W}_3 W_4 + W_4 \tilde{W}_3 W_2 \tilde{W}_1 \\
& - W_1 \tilde{W}_2 W_3 \tilde{W}_4 - \tilde{W}_4 W_3 \tilde{W}_2 W_1 \Big)
\end{split}
\end{equation}
and
\begin{equation}
\label{eq:Atr}
\begin{split}
\mathcal{A}_{\mathrm{tr}}[W]_{1234} &\\
= \hat{V}_{34}^2 \Big(
& + \big( \tilde{W}_1 W_2 + W_2 \tilde{W}_1 \big) \mathrm{tr}[\tilde{W}_3 W_4] \\
& - \big( W_1 \tilde{W}_2 + \tilde{W}_2 W_1 \big) \mathrm{tr}[W_3 \tilde{W}_4] \Big).
\end{split}
\end{equation}

As a remark, with this notation the conservative collision operator $\mathcal{C}_\mathrm{c}$ is of the form
\begin{multline}
\mathcal{C}_\mathrm{c}[W](k,t) = - \mathrm{i} \int_{\mathbb{T}^3} \ud k_2 \ud k_3 \ud k_4 \delta(\underline{k})\, \mathcal{P} \left(\tfrac{1}{\underline{\omega}}\right) \\
\times \big( \mathcal{A}[W]_{1234} - \mathcal{A}[W]_{1234}^* \big).
\end{multline}

\section{General properties of the Hubbard kinetic equation}
\label{sec:Properties}

The $\mathrm{SU}(2n+1)$ invariance of $H$ is reflected by
\begin{equation}
\mathcal{C}[U^* W U] = U^* \mathcal{C}[W] U	
\end{equation}
for all $U \in \mathrm{SU}(2n+1)$. Hence if $W(k,t)$ is a solution to the Boltzmann equation~\eqref{eq:BoltzmannEquation}, so is $U^*\,W(k,t)\,U$. Analogous to the Fermi case, hermiticity and positivity, $W(t) \geq 0$, is propagated in time. Positivity is enforced by the ``gain term'' in Eq.~\eqref{eq:AW1234def}.

In general, spin,
\begin{equation}
\label{eq:SpinConservation}
\int_{\mathbb{T}} \ud k \, W(k,t),
\end{equation}
and energy,
\begin{equation}
\label{eq:EnergyConservation}
\int_{\mathbb{T}} \ud k \, \omega(k) \, \mathrm{tr}[W(k,t)],
\end{equation}
are conserved. As discussed in \cite{BoltzmannFermi2012, BoltzmannNonintegrable2013}, additional conservation laws emerge depending on the dispersion relation $\omega(k)$. Namely, for the nearest neighbor hopping model, $\eta = 0$ in Eq.~\eqref{eq:omega_nnn}, the function
\begin{equation}
\label{eq:TraceConservation}
h(k,t) = \mathrm{tr}[W(k,t)] - \mathrm{tr}[W(\tfrac12-k,t)]
\end{equation}
remains constant in time (pointwise for each $k \in \mathbb{T}$). Using similar arguments as in the fermionic case, the conservation laws follow by an appropriate interchange of the integration variables $k_1,\dots,k_4$.

To prove the H-theorem, we first recall the definition of the entropy for bosons:
\begin{equation}
S[W] = \int_{\mathbb{T}} \ud k_1 \big( \mathrm{tr}[\tilde{W}_1 \log \tilde{W}_1] - \mathrm{tr}[W_1 \log W_1] \big).
\end{equation}
Hence the entropy production is given by
\begin{equation}
\label{eq:sigmaW}
\sigma[W] = \frac{\ud}{\ud t} S[W] = \int_{\mathbb{T}} \ud k_1 \, \mathrm{tr}[ (\log \tilde{W}_1 - \log W_1) \, \mathcal{C}[W]_1 ].
\end{equation}
The H-theorem states that
\begin{equation}
\label{eq:HTheorem}
\sigma[W] \geq 0 \qquad \text{for all positive semidefinite } W.
\end{equation}
To prove~\eqref{eq:HTheorem}, we start from the eigendecomposition (at fixed $t$)
\begin{equation}
\label{eq:WEigenDecomp}
W(k) = \sum_{\sigma} \lambda_\sigma(k) P_\sigma(k)	
\end{equation}
with eigenvalues $\lambda_\sigma(k) \ge 0$ and orthogonal eigen-projections $P_\sigma(k) = \lvert k,\sigma \rangle \langle k,\sigma\rvert$, such that $\langle k,\sigma | k, \sigma' \rangle = \delta_{\sigma \sigma'}$. As before, we use the notation $P_j = P_{\sigma_j}(k_j)$, $\lambda_j = \lambda_{\sigma_j}(k_j)$ and $\sum_{\boldsymbol{\sigma}} = \sum_{\sigma_1, \sigma_2, \sigma_3, \sigma_4}$. Inserting~\eqref{eq:WEigenDecomp} into~\eqref{eq:sigmaW} and using the representation in Eqs.~\eqref{eq:Aquad} and \eqref{eq:Atr} as well as the interchangeability $k_2 \leftrightarrow k_4$, one obtains
\begin{equation}
\begin{split}
\sigma[W] &= \pi \int_{\mathbb{T}^4} \ud^4 \boldsymbol{k} \, \delta(\underline{k}) \delta(\underline{\omega}) \\
&\ \times \sum_{\boldsymbol{\sigma}} \big( \log\tilde{\lambda}_1 - \log\lambda_1 \big) \big( \tilde{\lambda}_1 \lambda_2 \tilde{\lambda}_3 \lambda_4 - \lambda_1 \tilde{\lambda}_2 \lambda_3 \tilde{\lambda}_4 \big) \\
&\ \times \Big( \hat{V}_{34}^2 \, \mathrm{tr}[P_1 P_2] \mathrm{tr}[P_3 P_4] + \hat{V}_{23}^2 \, \mathrm{tr}[P_1 P_4] \mathrm{tr}[P_2 P_3] \\
&\quad + \hat{V}_{23}\hat{V}_{34} \, \mathrm{tr}[P_1 P_2 P_3 P_4] + \hat{V}_{23}\hat{V}_{34} \,\mathrm{tr}[P_4 P_3 P_2 P_1] \Big)\\
&= \pi \int_{\mathbb{T}^4} \ud^4 \boldsymbol{k} \, \delta(\underline{k}) \delta(\underline{\omega}) \sum_{\boldsymbol{\sigma}} \big( \tilde{\lambda}_1 \lambda_2 \tilde{\lambda}_3 \lambda_4 - \lambda_1 \tilde{\lambda}_2 \lambda_3 \tilde{\lambda}_4 \big)\\
&\ \times \log(\tilde{\lambda}_1/\lambda_1)\, \big\lvert \hat{V}_{34}\, \langle k_1,\sigma_1 | k_2,\sigma_2 \rangle \langle k_3,\sigma_3 | k_4,\sigma_4 \rangle \\
&\hspace{55pt} + \hat{V}_{23}\, \langle k_1,\sigma_1 | k_4,\sigma_4 \rangle \langle k_3,\sigma_3 | k_2,\sigma_2 \rangle \big\rvert^2.
\end{split}
\end{equation}
Interchanging $1 \leftrightarrow 3$, $2 \leftrightarrow 4$ and $(1,3) \leftrightarrow (2,4)$ and using $\hat{V}_{34} = \hat{V}_{12}$, $\hat{V}_{23} = \hat{V}_{14}$ due to $\delta(\underline{k})$, one arrives at
\begin{equation}
\label{eq:sigmaWfactorized}
\begin{split}
\sigma[W]
&= \frac{\pi}{4} \int_{\mathbb{T}^4} \ud^4 \boldsymbol{k} \, \delta(\underline{k}) \delta(\underline{\omega}) \\
&\ \times \sum_{\boldsymbol{\sigma}} \big( \tilde{\lambda}_1 \lambda_2 \tilde{\lambda}_3 \lambda_4 - \lambda_1 \tilde{\lambda}_2 \lambda_3 \tilde{\lambda}_4 \big) \log\!\bigg( \frac{\tilde{\lambda}_1 \lambda_2 \tilde{\lambda}_3 \lambda_4}{\lambda_1 \tilde{\lambda}_2 \lambda_3 \tilde{\lambda}_4} \bigg) \\
&\ \times \big\lvert \hat{V}_{34}\, \langle k_1,\sigma_1 | k_2,\sigma_2 \rangle \langle k_3,\sigma_3 | k_4,\sigma_4 \rangle \\
&\hspace{8pt} + \hat{V}_{23}\, \langle k_1,\sigma_1 | k_4,\sigma_4 \rangle \langle k_3,\sigma_3 | k_2,\sigma_2 \rangle \big\rvert^2.
\end{split}
\end{equation}
The last expression is $\ge 0$ since $(x-y) \log(x/y) \geq 0$.

From the form of~\eqref{eq:sigmaWfactorized} one concludes that the stationary states (discussed below) do not depend on the potential, as long as $\hat{V}(k)$ stays non-zero for all $k \in \mathbb{T}$.

\section{Stationary solutions}
\label{sec:StationarySolutions}

The kinematically allowed collisions depend only on the dispersion $\omega(k)$ and are discussed already in \cite{BoltzmannNonintegrable2013}.

The initial state determines a special, $k$-independent basis $\lvert\sigma\rangle$ through
\begin{equation}
\label{eq:ConservedSigmaBasis}
\int_{\mathbb{T}} \ud k \, W(k) = \sum_{\sigma} \varepsilon_\sigma \, \lvert\sigma \rangle \langle \sigma\rvert.
\end{equation}
By the spin conservation~\eqref{eq:SpinConservation} this basis is preserved in time. Thus it is natural to expand $W(k,t)$ in this special basis.

For long times, $W(k,t)$ will become diagonal in the conserved spin basis. Without the additional conservation laws in Eq.~\eqref{eq:TraceConservation}, $W(k,t)$ will converge to a thermal Bose-Einstein distribution
\begin{equation}
\label{eq:ThermalBoseEinstein}
W_\mathrm{th}(k) = \sum_{\sigma} \left( \mathrm{e}^{\beta(\omega(k) - \mu_\sigma)} - 1 \right)^{-1} \lvert\sigma\rangle\langle\sigma\rvert,
\end{equation}
with temperature $1/\beta$ and chemical potentials $\mu_\sigma$, precisely in accordance with the conserved spin and energy. For the nearest neighbor case with conserved $h(k,t)$, the stationary solutions have the same structure as in Eq.~\eqref{eq:ThermalBoseEinstein}, but with $\omega(k)$ replaced by a more general function $f$. One obtains
\begin{equation}
\label{eq:StationarySolutions}
W_{\mathrm{st}}(k) = \sum_{\sigma} \lambda_\sigma(k) \, \lvert\sigma\rangle\langle\sigma\rvert, \quad \lambda_\sigma(k) = \left( \mathrm{e}^{f(k) - a_\sigma} - 1 \right)^{-1},
\end{equation}
where $f$ is a real-valued, $1$-periodic function satisfying $f(k) = -f(\tfrac{1}{2} - k)$ and $f(k) - a_\sigma > 0$ for all $k$, $\sigma$.

Assuming that the initial $W$ converges to a stationary state of the form~\eqref{eq:StationarySolutions}, it must hold that
\begin{equation}
\label{eq:TraceDiffStationary}
h(k) = \sum_{\sigma} \Big( \big( \mathrm{e}^{f(k) - a_\sigma} - 1 \big)^{-1} - \big( \mathrm{e}^{- f(k) - a_\sigma} - 1 \big)^{-1} \Big).
\end{equation}

The spin conservation law requires that the eigenvalues $\varepsilon_\sigma$ in~\eqref{eq:ConservedSigmaBasis} are equal to
\begin{equation}
\label{eq:SpinStationary}
\varepsilon_\sigma = \int_\mathbb{T} \ud k\, \left( \mathrm{e}^{f(k) - a_\sigma} - 1 \right)^{-1}.
\end{equation}
We claim that~\eqref{eq:TraceDiffStationary} and~\eqref{eq:SpinStationary} uniquely determine $f$ and $a_\sigma$, or more specifically, that the map between
\begin{equation} \label{eq:SET1}
\mathrm{tr}[W(k)] - \mathrm{tr}[W(\tfrac12 - k)], \lvert k \rvert \le \tfrac14, \qquad \varepsilon_\sigma \geq 0 \text{ for all } \sigma
\end{equation}
and
\begin{multline} \label{eq:SET2}
\left\{f(k), a_\sigma \right\} \text{ with }
f(k) = -f(\tfrac{1}{2} - k) \text{ for } \abs{k} \le \tfrac14,\\
f(k) - a_\sigma > 0 \quad \text{ for all } k, \sigma
\end{multline}
is one-to-one. In particular, to a given $W$ one can associate a unique $W_\mathrm{st}$ of the form~\eqref{eq:StationarySolutions}.
\begin{proof}
By a short calculation,~\eqref{eq:TraceDiffStationary} can be written as
\begin{equation}
h(k) = \sum_{\sigma} \frac{-\sinh f(k)}{\cosh a_{\sigma} - \cosh f(k)}
\label{11}	
\end{equation}
and~\eqref{eq:SpinStationary} as
\begin{equation}
\varepsilon_\sigma = \int_\mathrm{I} \ud k \left( \frac{- \sinh a_\sigma }{\cosh a_\sigma - \cosh f(k)} - 1 \right)
\end{equation}
with interval of integration $\mathrm{I} = [-\tfrac14,\tfrac14 ]$. We define a generalized ``free energy'' through
\begin{equation}
H(f,a_{\sigma}) = \int_\mathrm{I} \ud k \sum_{\sigma} -\log\big(\cosh a_\sigma - \cosh f(k)\big).
\end{equation}
The map $(f,a_\sigma) \mapsto H$ is strictly convex: namely, $H$ is an integral and sum of functions
\begin{equation}
(f,a) \mapsto -\log(\cosh a - \cosh f), \quad \abs{f} < \abs{a}
\end{equation}
which are strictly convex since the eigenvalues of the Hessian matrix are $\cosh(a \pm f) - 1 > 0$. Furthermore
\begin{equation}
\frac{\partial}{\partial a_\sigma} H = \int_\mathrm{I} \ud k \frac{-\sinh a_\sigma }{\cosh a_\sigma - \cosh f(k) } = \varepsilon_\sigma - \frac{1}{2}	
\end{equation}
and
\begin{equation}
\frac{\delta H}{\delta f(k)} = \sum_{\sigma} \frac{\sinh f(k) }{\cosh a_\sigma - \cosh f(k) } = - h(k).
\end{equation}
Thus the map from above can be viewed as Legendre transform from the first set~\eqref{eq:SET1} to the second set of variables~\eqref{eq:SET2}. Since $H$ is convex, the map is one-to-one.
\end{proof}

\section{Derivation of the Boltzmann equation from the Bose-Hubbard hamiltonian}
\label{sec:Derivation}

We transcribe \cite{DerivationBoltzmann2013} to bosons and generalize to arbitrary (integer) spin quantum numbers. Notably, the determinants for fermions will be replaced by permanents for bosons in Eq.~\eqref{eq:WickRule} below, due to the switch from anticommutators to commutators. In addition, for this section we consider the straightforward generalization to $\mathbb{Z}^d$ as underlying lattice.

We start from the hamiltonian in Eq.~\eqref{eq:BoseHubbardHamiltonian} and assume as in \cite{BoltzmannFermi2012, BoltzmannNonintegrable2013, DerivationBoltzmann2013} that the initial state is gauge invariant, invariant under translations, and quasi-free. It is thus completely determined by the two point function
\begin{equation}
\langle \hat{a}_\sigma(k)^* \hat{a}_\tau(\tilde{k}) \rangle = \delta(k -\tilde{k}) W_{\sigma \tau}(k,0), \quad \sigma, \tau \in S
\end{equation}
where $\langle \cdot \rangle$ denotes the average with respect to the initial state and $S \equiv \{-n, \dots, n\}$ enumerates spin quantum numbers. Averages of the form $\langle (a^*)^m a^n \rangle$ vanish unless $m = n$, and all other moments are determined by the Wick pairing rule.
As discussed in \cite{DerivationBoltzmann2013}, the quasi-free property is approximately maintained up to times of order $\lambda^{-2}$ for small $\lambda \ll 1$.

We expand the true two-point function $W_\lambda$, defined by the relation $\delta(k - \tilde{k}) W_\lambda(k,t)_{\sigma \tau} = \langle \hat{a}_\sigma(k,t)^* \hat{a}_\tau(\tilde{k},t) \rangle$, for fixed $t$ up to order $\lambda^2$ as
\begin{equation} \label{eq:InitialExpansion}
W_\lambda(k,t) = W^{(0)}(k) + \lambda W^{(1)}(k,t) + \lambda^2 W^{(2)}(k,t) + \mathcal{O}(\lambda^3),
\end{equation}
and will extract the collision operator from $W^{(2)}$. To avoid a specific spin basis, choose arbitrary vectors $\mathrm{f}, \mathrm{g} \in \mathbb{C}^{2n+1}$ and consider $\langle \mathrm{f}, \, W_\lambda(k,t) \mathrm{g} \rangle$ where $\langle \cdot, \, \cdot \rangle$ denotes the inner product (anti-linear on the left) in spin space. We will use the vector valued operators
\begin{equation} \label{op:vec}
\hat{a}_\mathrm{f}(k)^* = \sum_{\sigma \in S} \overline{\mathrm{f}}_{\sigma} \, \hat{a}_{\sigma}(k)^* \, \mathfrak{e}_{\sigma} \text{ and }
\hat{a}_\mathrm{g}(k) = \sum_{\sigma \in S} \mathrm{g}_{\sigma} \, \hat{a}_{\sigma}(k) \, \mathfrak{e}_{\sigma},
\end{equation}
where $\overline{\mathrm{f}}$ denotes the complex conjugate, $\mathrm{f}_\sigma, \mathrm{g}_\sigma$, $\sigma \in S$ denote the components of $\mathrm{f}$ and $\mathrm{g}$ and $\mathfrak{e}_{\sigma}$ enumerates the standard basis. The following operations map two $(2n+1)$-vector valued operators into a scalar-valued one:
\begin{equation}
v \odot w = \sum_{\sigma, \tau \in S} v_\sigma w_\tau \quad \mathrm{and} \quad v \cdot w = \sum_{\sigma \in S} v_\sigma w_\sigma.	
\end{equation}
For instance,
\begin{equation*}
\langle \hat{a}_\mathrm{f}(k,t)^* \odot \hat{a}_\mathrm{g}(\tilde{k},t) \rangle = \delta(k - \tilde{k}) \, \langle \mathrm{f}, \, W_\lambda(k,t) \mathrm{g} \rangle.
\end{equation*}

The time derivative of the basic $(2n+1)$-vector valued operator becomes
\begin{multline}
\frac{\ud}{\ud t} \hat{a}_\mathrm{f}(k,t)^{\#} = \mathrm{i} [\hat{H}, \hat{a}_\mathrm{f}(k,t)^{\#}] \\
= \mathrm{i} [\hat{H}_0, \hat{a}_\mathrm{f}(k)^{\#}](t) + \mathrm{i}\, \frac{\lambda}{2}\, [\hat{H}_1, \hat{a}_\mathrm{f}(k)^{\#}](t)
\end{multline}
where $\#$ denotes either nothing or an adjoint (annihilation or creation operator). For the quadratic $H_0$ it follows directly from the commutation relations that
\begin{equation}
\begin{split}
\big[\hat{H}_0, \hat{a}_\mathrm{g}(k)\big] &= \int_{\mathbb{T}^d} \ud k' \, \omega(k') \big[\hat{a}(k')^* \cdot \hat{a}(k'), \hat{a}_\mathrm{g}(k)\big] \\
		&= - \omega(k) \, \hat{a}_\mathrm{g}(k)	
\end{split}
\end{equation}
and for the creation operator
\begin{equation}
[\hat{H}_0, \hat{a}_\mathrm{f}(k)^*] = - [H_0, \hat{a}_\mathrm{f}(k)]^* = \omega(k) \, \hat{a}_\mathrm{f}(k)^*.
\end{equation}
For $H_1$ we first consider
\begin{multline}
[H_1, a_\mathrm{g}(z)] = \frac{1}{2} \sum_{x \in \mathbb{Z}^d} V(x - z) \big( a(x)^* \cdot a(x) \big) \, a_\mathrm{g}(z) \\
	+ \frac{1}{2} \sum_{x \in \mathbb{Z}} V(z - x) a_\mathrm{g}(z) \, \big( a(x)^* \cdot a(x) \big)
\end{multline}
such that in momentum space
\begin{multline}
[\hat{H}_1, \hat{a}_\mathrm{g}(k_1)] = \sum_{z \in \mathbb{Z}^d} [H_1, a_\mathrm{g}(z)] \mathrm{e}^{-2 \pi \mathrm{i} \, k_1 \cdot z} \\
 = \frac{1}{2} \int_{\mathbb{T}^d} \ud k \hat{V}(k - k_1) \hat{a}_\mathrm{g}(k_1) 
 - \int_{(\mathbb{T}^d)^3} \ud k_{234} \, \delta(\underline{k}) \\
 \times \hat{V}(k_3 - k_4) 
	\hat{a}_\mathrm{g}(k_2) \big( \hat{a}(k_3)^* \cdot \hat{a}(k_4) \big).
\end{multline}
Thereby we obtain
\begin{equation}
\label{eq:evol}
\begin{split}
\frac{\ud}{\ud t} \hat{a}_\mathrm{g}(k,t) &= \mathrm{i} \, [\hat{H}, \hat{a}_\mathrm{g}(k_1,t)] \\
	&= - \mathrm{i} \, \omega(k) \, \hat{a}_\mathrm{g}(k_1,t) + \mathrm{i}\,\frac{\lambda}{2} \, V(0) \, \hat{a}_\mathrm{g}(k_1,t) \\
	&\quad - \mathrm{i} \, \lambda \int_{(\mathbb{T}^d)^3} \ud k_{234} \, \delta(\underline{k}) \, \hat{V}(k_3 - k_4) \\
	&\hspace{40pt} \times \hat{a}_\mathrm{g}(k_2,t) \, \big( \hat{a}(k_3,t)^* \cdot \hat{a}(k_4,t) \big) 
\end{split}
\end{equation}
where $\underline{k} = k_1 - k_2 + k_3 - k_4$. For the subsequent calculations, we use the notation $k_{1234} = (k_1,k_2,k_3,k_4)$ and introduce the following terms:
\begin{multline}
\mathcal{A}[h,a,b,c](k_1,t)
	= \int_{(\mathbb{T}^d)^3} \ud k_{234} \, \delta(\underline{k}) \, h(k_{1234},t) \\
		\times \hat{V}(k_3 - k_4) \, a(k_2,t) \, \big( b(k_3,t) \cdot c(k_4,t) \big)
\end{multline}
and
\begin{multline}
\mathcal{A}_*[\overline{h},a,b,c](k_1,t)
	= \int_{(\mathbb{T}^d)^3} \ud k_{234} \, \delta(\underline{k}) \, \overline{h}(k_{1234},t) \\
		\times \hat{V}(k_2 - k_3) \, \big( a(k_2,t) \cdot b(k_3,t) \big) \, c(k_4,t),
\end{multline}
where $h$ is any complex-valued function and $a, b, c$ are $(2n+1)$-component vector-valued operators as in \eqref{op:vec}. Then $\mathcal{A}$ and $\mathcal{A}_*$ are again vector-valued operators and satisfy the relation
\begin{equation}
\big( \mathcal{A}[h,a,b^*,c](k,t) \big)^* = \mathcal{A}_*[\overline{h},c^*,b,a^*](k,t).
\end{equation}
The evolution equation~\eqref{eq:evol} can then be written as
\begin{equation}
\begin{split}
\frac{\ud}{\ud t} \hat{a}_\mathrm{g}(k,t)
	&= - \mathrm{i} \big( \omega(k) - \tfrac{1}{2} \lambda \, V(0) \big) \, \hat{a}_\mathrm{g}(k,t) \\
		&\quad - \mathrm{i} \lambda \, \mathcal{A}[\mathrm{id},\hat{a}_\mathrm{g},\hat{a}^*,\hat{a}](k,t)
\end{split}
\end{equation}
and correspondingly for the creation operator
\begin{equation}
\begin{split}
\Big( \frac{\ud}{\ud t} \hat{a}_\mathrm{f}(k,t) \Big)^*
	&= \frac{\ud}{\ud t} \hat{a}_\mathrm{f}(k,t)^* \\
	&= \mathrm{i} \big( \omega(k) -\tfrac{1}{2} \lambda \, V(0) \big) \, \hat{a}_\mathrm{f}(k,t)^* \\
		&\quad + \mathrm{i} \lambda \, \mathcal{A}_*[\mathrm{id},\hat{a}^*,\hat{a},\hat{a}_\mathrm{f}^*](k,t).
\end{split}
\end{equation}
The linear part can be removed by defining
\begin{equation}
\mathfrak{a}_\mathrm{g}(k,t) = \mathrm{e}^{ \mathrm{i} ( \omega(k) - \tfrac{1}{2} \lambda {V}(0) ) t} \, \hat{a}_\mathrm{g}(k,t).
\end{equation}
%
%
The phase factor cancels in the correlator, such that
\begin{equation}
\langle \mathfrak{a}_\mathrm{f}(k,t)^* \odot \mathfrak{a}_\mathrm{g}(\tilde{k},t) \rangle
	= \langle \hat{a}_\mathrm{f}(k,t)^* \odot \hat{a}_\mathrm{g}(\tilde{k},t) \rangle.
\end{equation}
With the notation
\begin{equation}
\omega_{abcd} = \omega(k_a) - \omega(k_b) + \omega(k_c)  - \omega(k_d)   
\end{equation}
one finally arrives at
\begin{equation}
\label{eq:dt_ag}
\frac{\ud}{\ud t} \mathfrak{a}_\mathrm{g}(k_1,t)
	= - \mathrm{i} \lambda \, \mathcal{A}[\mathrm{e}^{\mathrm{i} \omega_{1234} t}, \mathfrak{a}_\mathrm{g}, \mathfrak{a}^*, \mathfrak{a}](k_1,t),
\end{equation}
and for the adjoint
\begin{equation}
\frac{\ud}{\ud t} \mathfrak{a}_\mathrm{f}(k_1,t)^*
	= \mathrm{i} \lambda \, \mathcal{A}_*[\mathrm{e}^{- \mathrm{i} \omega_{1234} t}, \mathfrak{a}^*, \mathfrak{a}, \mathfrak{a}_\mathrm{f}^*](k_1,t).
\end{equation}
Integrating Eq.~\eqref{eq:dt_ag} leads to
\begin{equation}
\mathfrak{a}_\mathrm{g}(k_1,t) = \mathfrak{a}_\mathrm{g}(k_1,0) - \mathrm{i} \lambda \int_{0}^t \ud s \, 
		\mathcal{A}[\mathrm{e}^{\mathrm{i} \omega_{1234} s}, \mathfrak{a}_\mathrm{g}, \mathfrak{a}^*, \mathfrak{a}](k_1,s),
\end{equation}
We now iterate Eq.~\eqref{eq:dt_ag} twice up to second order of the Dyson expansion, such that with an error of order $\lambda^3$
\begin{equation}
\begin{split}
&\frac{\ud}{\ud t} \mathfrak{a}_\mathrm{g}(k_1,t)
	= - \mathrm{i} \lambda \, \mathcal{A}[\mathrm{e}^{\mathrm{i} \omega_{1234} t}, \hat{a}_\mathrm{g}, \hat{a}^*, \hat{a}](k_1,0) \\
	&- \lambda^2 \, \int_0^t \ud s \,
		\mathcal{A}[\mathrm{e}^{\mathrm{i} \omega_{1234} t},
		\mathcal{A}[\mathrm{e}^{\mathrm{i} \omega_{2678} s}, \hat{a}_\mathrm{g}, \hat{a}^*, \hat{a}],
		\hat{a}^*, \hat{a}](k_1,s) \\
	&+ \lambda^2 \, \int_0^t \ud s \,
		\mathcal{A}[\mathrm{e}^{\mathrm{i} \omega_{1234} t}, \hat{a}_\mathrm{g}, 
		\mathcal{A}^*[\mathrm{e}^{-\mathrm{i} \omega_{3678} s}, \hat{a}^*, \hat{a}, \hat{a}^*], \hat{a}](k_1,s) \\
	&- \lambda^2 \, \int_0^t \ud s \,
		\mathcal{A}[\mathrm{e}^{\mathrm{i} \omega_{1234} t}, \hat{a}_\mathrm{g}, \hat{a}^*,
		\mathcal{A}[\mathrm{e}^{\mathrm{i} \omega_{4678} s}, \hat{a}, \hat{a}^*, \hat{a}]](k_1,s) \\
	&= \lambda \frac{\ud}{\ud t} \mathfrak{a}^{(1)}_\mathrm{g}(k_1,t) + \lambda^2 \frac{\ud}{\ud t} \mathfrak{a}^{(2)}_\mathrm{g}(k_1,t) + \mathcal{O}(\lambda^3).
\end{split}
\end{equation}
We have thus obtained the expansion in $\lambda$ (for fixed $t$)
\begin{equation}
\mathfrak{a}_\mathrm{g}(k_1,t) = \mathfrak{a}^{(0)}_\mathrm{g}(k_1,t) + \lambda \, \mathfrak{a}^{(1)}_\mathrm{g}(k_1,t) + \lambda^2 \, \mathfrak{a}^{(2)}_\mathrm{g}(k_1,t) + \mathcal{O}(\lambda^3),
\end{equation}
where $\mathfrak{a}^{(0)}_{\mathrm{g}}(k,t) = \mathfrak{a}^{(0)}_{\mathrm{g}}(k,0) =\hat{a}_\mathrm{g}(k)$. A corresponding expression is satisfied by $\mathfrak{a}_{\mathrm{f}}(k,t)^*$. Iterating further yields the formal expansion
\begin{multline} \label{expansion}
\frac{\ud}{\ud t} \langle \mathfrak{a}_\mathrm{f}(k,t)^* \odot \mathfrak{a}_\mathrm{g}(\tilde{k},t) \rangle \\
	 = \sum_{n=0}^\infty \lambda^n \sum_{m=0}^n \frac{\ud}{\ud t} \langle {\mathfrak{a}_\mathrm{f}(k,t)^*}^{(m)} \odot \mathfrak{a}_\mathrm{g}(\tilde{k},t)^{(n-m)} \rangle.
\end{multline}
Therefore, $W_\lambda(k,t)$ can be written as
\begin{multline} \label{eq:expansion2}
\delta(k - \tilde{k}) \, \langle \mathrm{f}, \, W_\lambda(k,t) \mathrm{g} \rangle = 
	\langle \mathfrak{a}_\mathrm{f}(k,0)^* \odot \mathfrak{a}_\mathrm{g}(\tilde{k},0) \rangle \\
+ \sum_{n=1}^\infty \lambda^n \int_0^t \ud s \sum_{m=0}^n \frac{\ud}{\ud s} 
		\langle {\mathfrak{a}_\mathrm{f}(k,s)^*}^{(m)} \odot \mathfrak{a}_\mathrm{g}(\tilde{k},s)^{(n-m)} \rangle \\
= \delta(k - \tilde{k}) \sum_{n=0}^\infty \lambda^n \langle \mathrm{f}, \, W^{(n)}(k,t) \mathrm{g} \rangle.
\end{multline}
The zeroth order term of Eq.~\eqref{eq:expansion2} reads
\begin{equation}
\begin{split}
\delta(k - \tilde{k}) \, \langle \mathrm{f}, \, W^{(0)}(k) \mathrm{g} \rangle 
	&= \langle \mathfrak{a}_\mathrm{f}(k,0)^* \odot \mathfrak{a}_\mathrm{g}(\tilde{k},0) \rangle \\
	&= \langle \hat{a}_\mathrm{f}(k)^* \odot \hat{a}_\mathrm{g}(\tilde{k}) \rangle.
\end{split}
\end{equation}
In the next two sections we compute the first and second order terms.

\subsection{First-order terms}

We represent the various summands of the $W^{(1)}(k,t)$ term in Eq.~\eqref{eq:expansion2} as Feynman diagrams, which coincide for fermions and bosons. The first order terms are determined by
\begin{equation}
\label{eq:full1storder}
\begin{split}
\delta(k_1 &- k_5) \langle \mathrm{f}, \, W^{(1)}(k_1,t) \mathrm{g} \rangle \\
&= \mathrm{i} \int_0^t \ud s \, \langle \mathcal{A}_*[\mathrm{e}^{-\mathrm{i} \omega_{1234} s}, \mathfrak{a}^*, \mathfrak{a}, \mathfrak{a}_\mathrm{f}^*](k_1) \odot \mathfrak{a}_{\mathrm{g}}(k_5,s)^{(0)} \rangle \\
&\ - \mathrm{i} \int_0^t \ud s \, \langle {\mathfrak{a}_{\mathrm{f}}(k_1,s)^*}^{(0)} \odot \mathcal{A}[\mathrm{e}^{\mathrm{i} \omega_{5234} s}, \mathfrak{a}_\mathrm{g}, \mathfrak{a}^*, \mathfrak{a}](k_5) \rangle \\
&= \mathrm{i} \int_0^t \ud s \int_{(\mathbb{T}^d)^3} \ud k_{234} \, \delta(\underline{k}) \, \hat{V}(k_2-k_3) \, \mathrm{e}^{-\mathrm{i} \omega_{1234} s} \\
&\qquad \times \langle \big( \hat{a}(k_2)^* \cdot \hat{a}(k_3) \big) 
		\big(\hat{a}_\mathrm{f}(k_4)^* \odot \hat{a}_{\mathrm{g}}(k_5) \big) \rangle \\
&\ - \mathrm{i} \int_0^t \ud s \int_{(\mathbb{T}^d)^3} \ud k_{234} \, \delta(\underline{k}) \, \hat{V}(k_3 - k_4) \, \mathrm{e}^{\mathrm{i} \omega_{5234} s} \\
&\qquad \times \langle \big( \hat{a}_\mathrm{f}(k_1)^* \odot \hat{a}_{\mathrm{g}}(k_2) \big) 
		\big(\hat{a}(k_3)^* \cdot \hat{a}(k_4) \big) \rangle.	
\end{split}
\end{equation}
\begin{figure*}[!ht]
\centering
\includegraphics[width=0.8\textwidth]{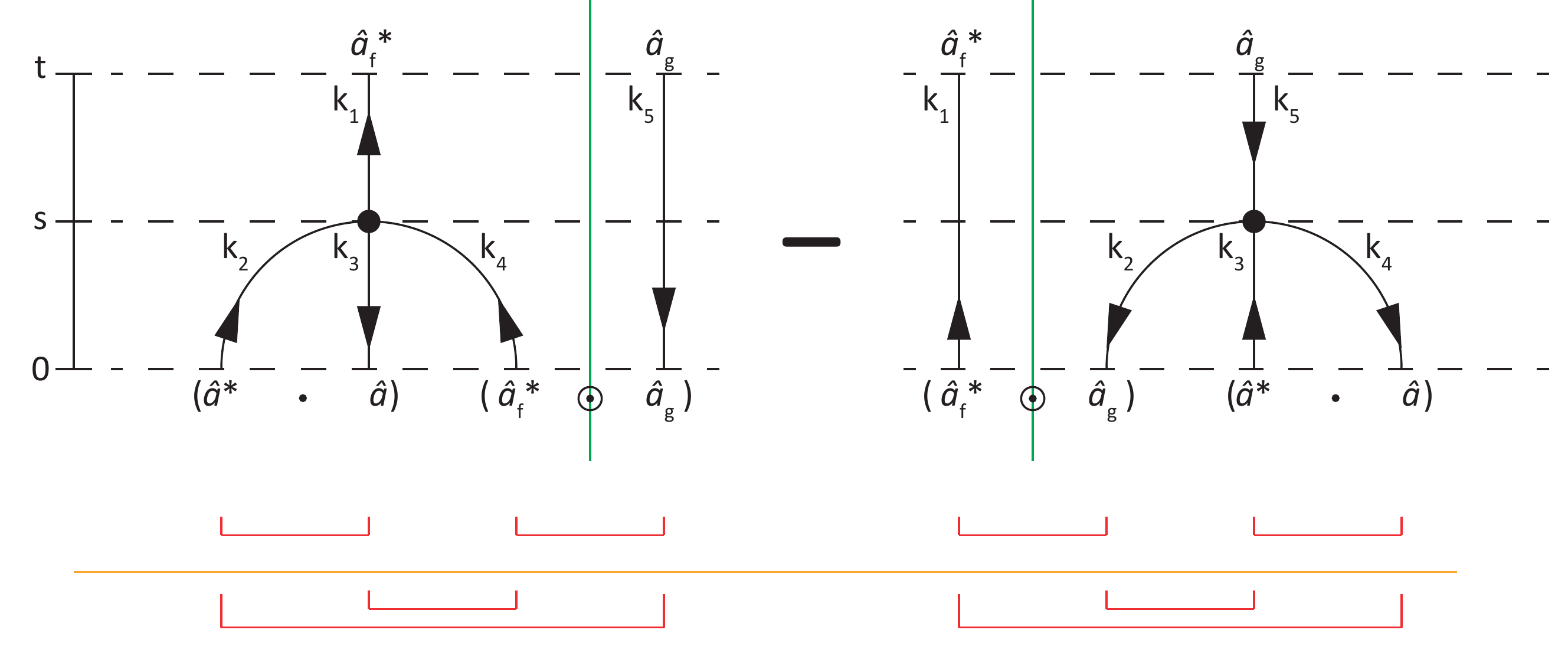}
\caption{The diagrams of the first order terms in $\lambda$.}
\label{fig:first_order_graphs}
\end{figure*}
The first term is represented by the left graph in Fig.~\ref{fig:first_order_graphs}. Each graph consists of the following symbols: vertices, edges and time slices. The time direction points from bottom to top. The $n$-th order terms have $n$ vertices, and so the first order terms have only a single vertex. The vertex represents the interaction of particles. The edges are labeled by oriented momentum-variables $k_i$. If the earlier of the endpoints is a creation operator, the arrow points in the time direction, and if it is an annihilation operator, the arrow points opposite to the time direction. Then, by definition of $\mathcal{A}$, at every vertex there are two ingoing and two outgoing arrows. 

To reconstruct the corresponding integral from a given graph, one needs to iteratively add the following five operations for each vertex:
\begin{enumerate}
	\item An integration of a time variable $s$ from zero to the end of the time slice after the vertex. In Fig.~\ref{fig:first_order_graphs} this amounts to using the time integral $\int_0^t \ud s$.
	\item The integration over the momentum variables can be read of as follows: one needs to add $\int_{(\mathbb{T}^d)^3} \ud k_{ijl}$ where  $k_i$, $k_j$ and $k_l$ label the three ``earlier'' edges.
	\item A product of four phase factors $\mathrm{e}^{\pm \mathrm{i} \omega(k_j) s}$, one for each arrow attached to the vertex, where $s$ denotes the time integration variable of the vertex. A negative sign is chosen if the arrow points in the time direction, and a positive sign if it points against the time direction. 
	\item A $\delta$-function ensuring the momentum conservation, in which a positive sign is used if the corresponding arrow points away from the vertex, and a negative sign if the arrow points towards the vertex. 
	\item A factor ``$\pm \mathrm{i}$'' with a positive sign if the single later edge points away from the vertex, and a negative sign if it points towards the vertex.
\end{enumerate}
Finally, the average $\langle \cdot \rangle$ of the product of creation and annihilation operators at the bottom of the graph needs to be taken. Every $(\hat{a}(k_i)^* \cdot \hat{a}(k_j))$ entails a factor of $\hat{V}(k_i - k_j)$. By construction, if one starts to count the direction of the arrows from left to right in any of the time slices, they always start with an up-arrow and alternate from left to right in up-down combinations. This results in an alternating sequence of creation and annihilation operators at the bottom of the graph. The Wick-pairings ``$\sqcup$'' shown under the graph follow from averaging this alternating sequence over the initial quasi-free state.  The average has a particularly simple form for the alternating order of creation and annihilation operators: it can then be computed according to the Wick rule
\begin{equation}
\label{eq:WickRule}
\langle \hat{a}_{i_1}^* \hat{a}_{j_1} \cdots \hat{a}_{i_n}^* \hat{a}_{j_n} \rangle = \mathrm{perm}[K(i_k,j_l)]_{1 \leq k,l \leq n}	
\end{equation}
where
\begin{equation}
K(i_k,j_l) = \begin{cases} \langle \hat{a}_{i_k}^* \hat{a}_{j_l} \rangle & \mathrm{if}\ \ k \leq l \\
	\langle \hat{a}_{j_l} \hat{a}_{i_k}^* \rangle & \mathrm{if}\ \ k > l \end{cases}
\end{equation}
and ``$\mathrm{perm}$'' denotes the permanent of a matrix. For instance, the expectation value $\langle \cdot \rangle$ over the initial state in the first term in Eq.~\eqref{eq:full1storder} can be expressed as
\begin{equation}
\label{wick:perm}
\begin{split}
&\langle \big( \hat{a}(k_2)^* \cdot \hat{a}(k_3) \big) \big(\hat{a}_\mathrm{f}(k_4)^* \odot \hat{a}_{\mathrm{g}}(k_5) \big) \rangle \\
&= \sum_{\sigma_1, \sigma, \tau \in S} \overline{\mathrm{f}}_\sigma \mathrm{g}_\tau 
		\langle \hat{a}_{\sigma_1}(k_2)^* \hat{a}_{\sigma_1}(k_3) \hat{a}_{\sigma}(k_4)^* \hat{a}_{\tau}(k_5) \rangle \\
&= \sum_{\sigma_1, \sigma, \tau \in S} \overline{\mathrm{f}}_\sigma \mathrm{g}_\tau \\
&\qquad \times \mathrm{perm}\!\begin{bmatrix}
	\langle \hat{a}_{\sigma_1}(k_2)^* \hat{a}_{\sigma_1}(k_3) \rangle &
	\langle \hat{a}_{\sigma_1}(k_2)^* \hat{a}_{\tau}(k_5) \rangle \\
	- \langle \hat{a}_{\sigma_1}(k_3) \hat{a}_{\sigma}(k_4)^* \rangle &
	\langle \hat{a}_{\sigma}(k_4)^* \hat{a}_{\tau}(k_5) \rangle
\end{bmatrix} \\
&= \sum_{\sigma_1, \sigma, \tau \in S} \overline{\mathrm{f}}_\sigma \mathrm{g}_\tau \big( \langle \hat{a}_{\sigma_1}(k_2)^* \hat{a}_{\sigma_1}(k_3) \rangle \langle \hat{a}_{\sigma}(k_4)^* \hat{a}_{\tau}(k_5) \rangle \\
&\qquad\qquad + \langle \hat{a}_{\sigma_1}(k_3) \hat{a}_{\sigma}(k_4)^* \rangle \langle \hat{a}_{\sigma_1}(k_2)^* \hat{a}_{\tau}(k_5) \rangle \big).
\end{split}
\end{equation}
The two Wick pairings shown in Fig.~\ref{fig:first_order_graphs} represent the two different pairings in equation \eqref{wick:perm}.  Since for instance, $\langle \hat{a}_{\sigma_1}(k_3) \hat{a}_\sigma(k_4)^* \rangle = \delta(k_3-k_4) \tilde{W}(k_4)_{\sigma\sigma_1}$, the left diagram yields
\begin{multline}
\int_0^t \ud s \, \langle {\dot{\mathfrak{a}}_{\mathrm{f}}(k_1,s)^*}^{(1)} \odot \mathfrak{a}_{\mathrm{g}}(k_5,s)^{(0)} \rangle = \mathrm{i} t \, \delta(k_1 - k_5) \\
\ \times \int_{\mathbb{T}^d} \ud k_2 \big( \hat{V}(0) \, \mathrm{tr}[W_2] \langle \mathrm{f}, W_1 \mathrm{g} \rangle + \hat{V}(k_1-k_2) \langle \mathrm{f}, \tilde{W}_2 W_1 \mathrm{g} \rangle \big)
\end{multline}
where $\dot{\mathfrak{a}}(k,t)=\frac{\ud}{\ud t} \mathfrak{a}(k,t)$. The contribution of the right diagram in Fig.~\ref{fig:first_order_graphs} can also be computed directly by taking an adjoint of the result above, yielding
\begin{multline}
\int_0^t \ud s \, \langle {\mathfrak{a}_{\mathrm{f}}(k_1,s)^*}^{(0)} \odot \dot{\mathfrak{a}}_{\mathrm{g}}(k_5,s)^{(1)} \rangle
= - \mathrm{i} t \, \delta(k_1 - k_5) \\
\times \int_{\mathbb{T}^d} \ud k_2 \big( \hat{V}(0) \, \mathrm{tr}[W_2] \langle \mathrm{f}, W_1 \mathrm{g} \rangle 
		+ \hat{V}(k_1 - k_2) \langle \mathrm{f}, W_1 \tilde{W}_2 \mathrm{g} \rangle \big).	
\end{multline}
Thus the first order term is given by
\begin{equation}
\begin{split}
W^{(1)}(k_1,t) &= - \mathrm{i} t [ R[W]_1, W_1 ], \\
	R[W]_1 &= \int_{\mathbb{T}^d} \ud k \, \hat{V}(k_1 - k) \, W(k) \in \mathbb{C}_{(2n+1) \times (2n+1)}.
\end{split}
\end{equation}
All four diagrams in Fig.~\ref{fig:first_order_graphs} have an interaction with zero momentum transfer (for instance, using the top left pairing leads to $k_4 = k_1$). Such diagrams will also appear in the second order and we call them \emph{zero momentum transfer diagrams}.

\subsection{Second-order terms}
We next consider the second order term in $\lambda$, which we decompose into a sum of four terms, obtained by evaluating the time-derivative in the equality
\begin{multline}
\label{W:2}
\delta(k - \tilde{k}) \langle \mathrm{f}, \, W^{(2)}(k,t) \mathrm{g} \rangle \\
= \int_0^t \ud s \sum_{m=0}^2 \frac{\ud}{\ud s} \big\langle {\mathfrak{a}_\mathrm{f}(k,s)^*}^{(m)} \odot \mathfrak{a}_\mathrm{g}(\tilde{k},s)^{(2-m)} \big\rangle.
\end{multline}

\subsection*{(1$\bf{'}$,1)-term}

In the previous section we have already shown that
\begin{multline}
\label{eq:11start}
\int_0^t \ud s \, \langle {\dot{\mathfrak{a}}_\mathrm{f}(k_1,s)^*}^{(1)} \odot \mathfrak{a}_\mathrm{g}(k_5,s)^{(1)} \rangle \\
= \int_0^t \ud s_2 \int_{0}^{s_2} \ud s_1 \, \langle \mathcal{A}^*[\mathrm{e}^{-\mathrm{i} \omega_{1234} s_2}, \mathfrak{a}^*, \mathfrak{a}, \mathfrak{a}_\mathrm{f}^*](k_1) \\
\odot \mathcal{A}[\mathrm{e}^{\mathrm{i} \omega_{5678} s_1}, \mathfrak{a}_\mathrm{g}, \mathfrak{a}^*, \mathfrak{a}](k_5) \rangle
\end{multline}
which can be represented by the Feynman diagram in Ref.~\cite[Fig.~2]{DerivationBoltzmann2013}. In order to evaluate the diagram we start with
\begin{multline}
\big\langle \big( a(k_2)^* \cdot a(k_3) \big) \big( a_\mathrm{f}(k_4)^* \odot a_\mathrm{g}(k_6) \big) \big( a(k_7)^* \cdot a(k_8) \big) \big\rangle \\
= \sum_{\sigma,\tau,\mu_1,\mu_2} \overline{\mathrm{f}}_\sigma \mathrm{g}_\tau \big\langle a_{\mu_1}(k_2)^* \, a_{\mu_1}(k_3) \\
\times a_\sigma(k_4)^* \, a_\tau(k_6) \, a_{\mu_2}(k_7)^* \, a_{\mu_2}(k_8) \big\rangle.
\end{multline}
Using Eq.~\eqref{eq:perm2} in appendix~\ref{sec:BosonicCorrelations},
\begin{multline}
\big\langle \hat{a}_{s_1}(i_1)^* \, \hat{a}_{r_1}(j_1) \, \hat{a}_{s_2}(i_2)^* \, \hat{a}_{r_2}(j_2) \, \hat{a}_{s_3}(i_3)^* \, \hat{a}_{r_3}(j_3) \big\rangle = \mathrm{perm} \\
\begin{bmatrix}
	\langle \hat{a}_{s_1}(i_1)^* \hat{a}_{r_1}(j_1) \rangle &
	\langle \hat{a}_{s_1}(i_1)^* \hat{a}_{r_2}(j_2) \rangle &
	\langle \hat{a}_{s_1}(i_1)^* \hat{a}_{r_3}(j_3) \rangle \\
	\langle \hat{a}_{r_1}(j_1) \hat{a}_{s_2}(i_2)^* \rangle &
	\langle \hat{a}_{s_2}(i_2)^* \hat{a}_{r_2}(j_2) \rangle &
	\langle \hat{a}_{s_2}(i_2)^* \hat{a}_{r_3}(j_3) \rangle \\
	\langle \hat{a}_{r_1}(j_1) \hat{a}_{s_3}(i_3)^* \rangle &
	\langle \hat{a}_{r_2}(j_2) \hat{a}_{s_3}(i_3)^* \rangle &
	\langle \hat{a}_{s_3}(i_3)^* \hat{a}_{r_3}(j_3) \rangle 
\end{bmatrix}
\end{multline}
one arrives at
\begin{equation}
\label{eq:11interm}
\begin{split}
&\langle \big( a(k_2)^* \cdot a(k_3) \big) \big( a_\mathrm{f}(k_4)^* \odot a_\mathrm{g}(k_6) \big) \big( a(k_7)^* \cdot a(k_8) \big) \rangle \\
	&= \delta(k_3 - k_7) \delta(k_4 - k_6) \delta(k_2 - k_8) \langle \mathrm{f}, \, W_4 \mathrm{tr}[\tilde{W}_3 W_2] \mathrm{g} \rangle \\
	&\ + \delta(k_2 - k_6) \delta(k_4 - k_8) \delta(k_3 - k_7) \langle \mathrm{f}, \, W_4 \tilde{W}_3 W_2 \mathrm{g} \rangle \\
	&\ + \delta(k_2 - k_3) \delta(k_4 - k_6) \delta(k_7 - k_8) \langle \mathrm{f}, \, W_4 \mathrm{tr}[W_2] \mathrm{tr}[W_7] \mathrm{g} \rangle \\
	&\ + \delta(k_2 - k_8) \delta(k_3 - k_4) \delta(k_6 - k_7) \langle \mathrm{f}, \, \tilde{W}_4 W_2 \tilde{W}_6 \mathrm{g} \rangle \\
	&\ + \delta(k_6 - k_7) \delta(k_4 - k_8) \delta(k_2 - k_3) \langle \mathrm{f}, \, W_4 \tilde{W}_6 \mathrm{tr}[W_2] \mathrm{g} \rangle \\
	&\ + \delta(k_7 - k_8) \delta(k_3 - k_4) \delta(k_2 - k_6) \langle \mathrm{f}, \, \tilde{W}_3 W_2 \mathrm{tr}[W_7] \mathrm{g} \rangle.
\end{split}
\end{equation}
Inserting this formula into \eqref{eq:11start} yields the following expression for the $(1',1)$-term,
\begin{equation}
\begin{split}
&\int_0^t \ud s \, \langle {\dot{\mathfrak{a}}_\mathrm{f}(k_1,s)^*}^{(1)} \odot \mathfrak{a}_\mathrm{g}(k_5,s)^{(1)} \rangle \\
	&= \delta(k_1 - k_5) \, \frac{1}{2} t^2 \, \langle \mathrm{f}, \, \mathcal{Z}[W]_1^{(1' 1)} \mathrm{g} \rangle \\
	&\quad + \delta(k_1 - k_5) \int_0^t \ud s_2 \int_0^{s_2} \ud s_1 \, 
		\int_{(\mathbb{T}^d)^3} \ud k_{234} \, \delta(\underline{k}) \\
	&\hspace{80pt} \times \mathrm{e}^{-\mathrm{i} \omega_{1234} (s_2-s_1)}
		\langle \mathrm{f}, \, \mathcal{D}[W]^*_{234} \, \mathrm{g} \rangle. 
\end{split}
\end{equation}
Here
\begin{multline}
\mathcal{D}[W]^*_{234} = V(k_2 - k_3)^2 \, W_4 \mathrm{tr}[\tilde{W}_3 W_2] \\
+ V(k_2 - k_3) V(k_3 - k_4) \, W_4 \tilde{W}_3 W_2  	
\end{multline}
and it results from the first two terms in Eq.~\eqref{eq:11interm}. The remaining four terms all leads to a diagram with a zero momentum transfer and summing up their contribution yields
\begin{multline}
\mathcal{Z}[W]_1^{(1' 1)}
= \hat{V}(0) \{ W_1, \, R[\tilde{W}]_1 \} \, \mathrm{tr}[R] \\
+ R[\tilde{W}]_1 \, W_1 \, R[\tilde{W}]_1 
+ \hat{V}(0)^2 \, W_1 \, \mathrm{tr}[R] \mathrm{tr}[R].	
\end{multline}

\subsection*{(1,1$\bf{'}$)-term}

A similar discussion applies to
\begin{multline}
\int_0^t \ud s \, \langle {\mathfrak{a}_\mathrm{f}(k_1,s)^*}^{(1)} \odot \dot{\mathfrak{a}}_\mathrm{g}(k_5,s)^{(1)} \rangle \\
\qquad = \int_0^t \ud s_2 \int_0^{s_2} \ud s_1 \,  
	\langle \mathcal{A}_*[\mathrm{e}^{-\mathrm{i} \omega_{1234} s_1}, \mathfrak{a}^*, \mathfrak{a}, \mathfrak{a}_\mathrm{f}^*](k_1) \\
	\odot \mathcal{A}[\mathrm{e}^{\mathrm{i} \omega_{5678} s_2}, \mathfrak{a}_\mathrm{g}, \mathfrak{a}^*, \mathfrak{a}](k_5) \rangle,
\end{multline}
which can also be computed by taking the adjoint of the $(1',1)$-term. This shows that 
\begin{equation}
\begin{split}
&\int_0^t \ud s \, \langle {\mathfrak{a}_\mathrm{f}(k_1,s)^*}^{(1)} \odot \dot{\mathfrak{a}}_\mathrm{g}(k_5,s)^{(1)} \rangle \\
	&= \delta(k_1 - k_5) \, \frac{1}{2} t^2 \, \langle \mathrm{f}, \, \mathcal{Z}[W]_1^{(1 1')} \mathrm{g} \rangle \\
&\qquad + \delta(k_1 - k_5) \int_0^t \ud s_2 \int_0^{s_2} \ud s_1 \int_{(\mathbb{T}^d)^3} \ud k_{234} \, \delta(\underline{k}) \\
&\hspace{80pt}\times \mathrm{e}^{\mathrm{i} \omega_{1234} (s_2-s_1)} \langle \mathrm{f}, \, \mathcal{D}[W]_{234} \mathrm{g} \rangle,
\end{split}
\end{equation}
where $\mathcal{Z}[W]_1^{(1 1')}= (\mathcal{Z}[W]_1^{(1' 1)})^* = \mathcal{Z}[W]_1^{(1' 1)} $ and
\begin{multline}
\mathcal{D}[W]_{234} = \hat{V}(k_2 - k_3)^2 \, W_4 \mathrm{tr}[\tilde{W}_3 W_2] \\
+ \hat{V}(k_2 - k_3) \hat{V}(k_3 - k_4) \, W_4 \tilde{W}_3 W_2,  	
\end{multline}
such that it hold $\mathcal{D}[W]^*_{234} = \mathcal{D}[W]_{234}$ by interchanging $k_2 \leftrightarrow k_4$ for the second term.

\subsection*{(2,0)-term}
The $(2,0)$-term is given by the following expression
\begin{equation}
\label{eq:20term}
\begin{split}
&\int_0^t \ud s \, \langle {\dot{\mathfrak{a}}_\mathrm{f}(k_1,s)^*}^{(2)} \odot \mathfrak{a}_\mathrm{g}(k_5,s)^{(0)} \rangle = \\
& - \int_0^t \ud s_2 \int_0^{s_2} \ud s_1 \, \langle \mathcal{A}_*[\mathrm{e}^{-\mathrm{i} \omega_{1234} s_2},\\
	&\hspace{40pt} \mathcal{A}_*[\mathrm{e}^{-\mathrm{i} \omega_{2678} s_1}, \mathfrak{a}^*, \mathfrak{a}, \mathfrak{a}^*], \mathfrak{a}, \mathfrak{a}_\mathrm{f}^*](k_1) \odot \mathfrak{a}_\mathrm{g}(k_5) \rangle \\
& + \int_0^t \ud s_2 \int_0^{s_2} \ud s_1 \, \langle \mathcal{A}_*[\mathrm{e}^{-\mathrm{i} \omega_{1234} s_2}, \mathfrak{a}^*, \\
	&\hspace{40pt} \mathcal{A}[\mathrm{e}^{\mathrm{i} \omega_{3678} s_1},
		\mathfrak{a}, \mathfrak{a}^*, \mathfrak{a}], \mathfrak{a}_\mathrm{f}^*](k_1) \odot \mathfrak{a}_\mathrm{g}(k_5) \rangle \\
& - \int_0^t \ud s_2 \int_0^{s_2} \ud s_1 \, \langle \mathcal{A}_*[\mathrm{e}^{-\mathrm{i} \omega_{1234} s_2}, \mathfrak{a}^*, \mathfrak{a}, \\
	&\hspace{40pt} \mathcal{A}_*[ \mathrm{e}^{-\mathrm{i} \omega_{4678} s_1}, \mathfrak{a}^*, \mathfrak{a}, \mathfrak{a}_\mathrm{f}^*]](k_1) 
		\odot \mathfrak{a}_\mathrm{g}(k_5) \rangle.
\end{split}
\end{equation}
To evaluate the contribution of the parings to the first term in Eq.~\eqref{eq:20term} we use
\begin{equation}
\begin{split}
&\langle \big( a(k_6)^* \cdot a(k_7) \big) \big( a(k_8)^* \cdot a(k_3) \big) \big( a_\mathrm{f}(k_4)^* \odot a_\mathrm{g}(k_5) \big) 
\rangle \\
&= \sum_{\sigma,\tau,\mu_1,\mu_2} \mathrm{f}_\sigma \mathrm{g}_\tau \langle a_{\mu_1}(k_6)^* a_{\mu_1}(k_7) a_{\mu_2}(k_8)^* a_{\mu_2}(k_3) a_\sigma(k_4)^* a_\tau(k_5) \rangle \\
& = \delta(k_7 - k_4) \delta(k_8 - k_3) \delta(k_6 - k_5) \, \langle \mathrm{f}, \, \tilde{W}_4 W_1 \mathrm{tr}[W_3] \mathrm{g} \rangle \\
&\quad + \delta(k_6 - k_3) \delta(k_8 - \tilde{k}) \delta(k_7 - k_4) \, \langle \mathrm{f}, \, \tilde{W}_4 W_3 W_1 \mathrm{g} \rangle \\
&\quad + \text{zero momentum transfer diagrams}
\end{split}
\end{equation}
and the contributions of the second term in Eq.~\eqref{eq:20term} are
\begin{equation}
\begin{split}
&\langle \big( a(k_2)^* \cdot a(k_6) \big) \big( a(k_7)^* \cdot a(k_8) \big) \big( a_\mathrm{f}(k_3)^* \odot a_\mathrm{g}(k_5) \big) \rangle \\
&= \sum_{\sigma,\tau,\mu_1,\mu_2} \mathrm{f}_\sigma \mathrm{g}_\tau \\
&\quad \times \langle a_{\mu_1}(k_2)^* a_{\mu_1}(k_6) a_{\mu_2}(k_7)^* a_{\mu_2}(k_8) a_\sigma(k_4)^* a_\tau(k_5) \rangle \\
&= \delta(k_8 - k_4) \delta(k_7 - \tilde{k}) \delta(k_6 - k_2) \, \langle \mathrm{f}, \, \tilde{W}_4 W_1 \mathrm{tr}[W_2] \mathrm{g} \rangle \\
&\quad + \delta(k_2 - k_8) \delta(k_7 - \tilde{k}) \delta(k_6 - k_4) \, \langle \mathrm{f}, \, \tilde{W}_4 W_2 W_1 \mathrm{g} \rangle \\
&\quad + \text{zero momentum transfer diagrams}
\end{split}
\end{equation}
and the contributions of the third term \eqref{eq:20term} are given by
\begin{equation}
\begin{split}
&\langle \big( a(k_2)^* \cdot a(k_3) \big) \big( a(k_6)^* \cdot a(k_7) \big) \big( a_\mathrm{f}(k_8)^* \odot a_\mathrm{g}(k_5) \big) 
\rangle \\
&= \sum_{\sigma,\tau,\mu_1,\mu_2} \mathrm{f}_\sigma \mathrm{g}_\tau \\
&\quad \times \langle a_{\mu_1}(k_2)^* a_{\mu_1}(k_3) a_{\mu_2}(k_6)^* a_{\mu_2}(k_7) a_\sigma(k_8)^* a_\tau(k_5) \rangle \\
&= \delta(k_8 - \tilde{k}) \delta(k_3 - k_6) \delta(k_2 - k_7) \, \langle \mathrm{f}, \, W_1 \mathrm{tr}[\tilde{W}_3 W_2 ] \mathrm{g} \rangle \\ 
	&\quad + \delta(k_2 - k_7) \delta(k_6 - \tilde{k}) \delta(k_3 - k_8) \, \langle \mathrm{f}, \, \tilde{W}_3 W_2 W_1 \mathrm{g} \rangle \\
	&\quad + \text{zero momentum transfer diagrams}.
\end{split}
\end{equation}
With the definitions
\begin{equation}
\begin{split}
&\mathcal{B}[W]^*_{1234} \\
&= \hat{V}(k_2 - k_3) \hat{V}(k_3 - k_4) \\
&\qquad \times \big( \tilde{W}_4 W_2 W_1 - \tilde{W}_4 W_3 W_1 - \tilde{W}_3 W_2 W_1 \big) \\
&\quad + V(k_2 - k_3)^2 \\
&\qquad \times \big( \tilde{W}_4 W_1 \mathrm{tr}[W_2] - \tilde{W}_4 W_1 \mathrm{tr}[W_3] - W_1 \mathrm{tr}[\tilde{W}_3 W_2] \big) 	
\end{split}
\end{equation}
and
\begin{multline}
\mathcal{Z}[W]_1^{(2 0)} = 
	- \hat{V}(0)^2 \, W_1 \, \mathrm{tr}[R] \mathrm{tr}[R] 
	- R[\tilde{W}]_1 \, R[\tilde{W}]_1 \, W_1 \\
	- \hat{V}(0) \, R[\tilde{W}]_1 \, W_1 \, \mathrm{tr}[R]
	- \hat{V}(0) \, R[\tilde{W}]_1 \, W_1 \, \mathrm{tr}[R]
\end{multline}
we obtain
\begin{equation}
\begin{split}
&\int_0^t \ud s \, \langle {\dot{\mathfrak{a}}_\mathrm{f}(k_1,t)^*}^{(2)} \odot \mathfrak{a}_\mathrm{g}(k_5,t)^{(0)} \rangle \\
&= \delta(k_1 - k_5) \frac{1}{2} t^2 \, \langle \mathrm{f}, \, \mathcal{Z}[W]_1^{(2 0)} \mathrm{g} \rangle \\
&\quad + \delta(k_1 - k_5) \int_0^t \ud s_1 \int_0^{s_1} \ud s_2 \int_{(\mathbb{T}^d)^3} \ud k_{234} \, \delta(\underline{k}) \\
&\hspace{70pt} \times \mathrm{e}^{-\mathrm{i} \omega_{1234} (s_2-s_1)} \langle \mathrm{f}, \, \mathcal{B}[W]^*_{1234} \, \mathrm{g} \rangle.
\end{split}
\end{equation}

\subsection*{(0,2)-term}

Analogous to the $(2,0)$-term, one arrives at
\begin{equation}
\begin{split}
&\int_0^t \ud s \, \langle {\mathfrak{a}_\mathrm{f}(k_1,s)^*}^{(0)} \odot \dot{\mathfrak{a}}_\mathrm{g}(k_5,s)^{(2)} \rangle \\
&= \delta(k_1 - k_5) \, \frac{1}{2} t^2 \, \langle \mathrm{f}, \, \mathcal{Z}[W]^{(0 2)} \mathrm{g} \rangle \\
&\quad + \delta(k_1 - k_5) \int_0^t \ud s_2 \int_0^{s_2} \ud s_1  \int_{(\mathbb{T}^d)^3} \ud ^3 k_{234}\, \delta(\underline{k}) \\
&\hspace{70pt} \times \mathrm{e}^{\mathrm{i} \omega_{1234} (s_2-s_1)} \langle \mathrm{f}, \, \mathcal{B}[W]_{1234} \, \mathrm{g} \rangle \, .
\end{split}
\end{equation}

\subsection{The limit $\lambda \rightarrow 0$, $t = \mathcal{O}(\lambda^{-2})$}
Before we consider the limit $\lambda \rightarrow 0$ we summarize all second order diagrams. Defining
\begin{align}
\mathcal{A}[W]_\mathrm{1234} &= \mathcal{D}[W]_\mathrm{234} + \mathcal{B}[W]_\mathrm{1234},\\
\mathcal{A}[W]^*_\mathrm{1234} &= \mathcal{D}[W]^*_\mathrm{234} + \mathcal{B}[W]^*_\mathrm{1234},
\end{align}
and using the identity
\begin{equation}
\begin{split}
&- [R[W]_1, \, [R[W]_1, \, W_1]] \\
&\quad = \mathcal{Z}[W]_1^{(1' 1)} + \mathcal{Z}[W]_1^{(1 1')} 
	+ \mathcal{Z}[W]_1^{(2 0)} + \mathcal{Z}[W]_1^{(0 2)},
\end{split}
\end{equation}
we thus find that 
\begin{equation}
\begin{split}
& \int_0^t \ud s \, \frac{\ud}{\ud s} \sum_{m=0}^2 \langle {\mathfrak{a}_\mathrm{f}(k_1,s)^*}^{(m)} \odot \mathfrak{a}_\mathrm{g}(k_5,s)^{(2-m)} \rangle \\ 
&= - \delta(k_1 - k_5) \, \frac{1}{2} t^2 \, \langle \mathrm{f}, \, [R[W]_1, \, [R[W]_1, \, W_1]] \mathrm{g} \rangle \\ 
&\quad + \delta(k_1 - k_5) \int_0^t \ud s_1 \int_0^{s_1} \ud s_2 \int_{(\mathbb{T}^d)^3} \ud k_{234} \, \delta(\underline{k}) \\
&\hspace{60pt} \times \mathrm{e}^{\mathrm{i} \omega_{1234} (s_2-s_1)} \langle \mathrm{f}, \, \mathcal{A}[W]_{1234} \, \mathrm{g} \rangle \\ 
&\quad + \delta(k_1 - k_5) \int_0^t \ud s_1 \int_0^{s_1} \ud s_2 \int_{(\mathbb{T}^d)^3} \ud k_{234} \, \delta(\underline{k}) \\
&\hspace{60pt} \times \mathrm{e}^{-\mathrm{i} \omega_{1234} (s_2-s_1)} \langle \mathrm{f}, \, \mathcal{A}[W]^*_{1234} \, \mathrm{g} \rangle .
\end{split}
\end{equation}
Hence the second order term $W^{(2)}$ is given by
\begin{equation}
W^{(2)}(k_1,t) = W^{(2)}_\mathrm{z}(k_1,t) + W^{(2)}_\mathrm{c}(k_1,t)
\end{equation}
where
\begin{equation}
\label{eq:sumZ2}
W^{(2)}_\mathrm{z}(k_1,t)= - \frac{1}{2} t^2 \, [R[W]_1, \, [R[W]_1, \, W_1]], 
\end{equation}
and
\begin{equation}
\label{eq:W2cdef}
\begin{split}
&W^{(2)}_\mathrm{c}(k_1,t)
= \int_0^t \ud s_1 \int_0^{s_1} \ud s_2 \int_{(\mathbb{T}^d)^3} \ud k_{234} \, \delta(\underline{k}) \\
&\times \big( \mathrm{e}^{\mathrm{i} \omega_{1234} (s_1-s_2)} \mathcal{A}[W]_{1234} + \mathrm{e}^{-\mathrm{i} \omega_{1234} (s_1-s_2)} \mathcal{A}[W]^*_{1234} \big).
\end{split}
\end{equation}
The collision operator is determined by taking at second order the limit $\lambda \rightarrow 0$ and simultaneous long times $\lambda^{-2} t$ with $t$ of order $1$. More explicitly,
\begin{equation}
t \, \mathcal{C}[W^{(0)}](k) = \lim_{\lambda \rightarrow 0} \lambda^2 \, W^{(2)}_\mathrm{c}(k, \lambda^{-2} t),
\end{equation}
where $W^{(2)}_\mathrm{c}$ is defined in \eqref{eq:W2cdef}.
To evaluate the limit, we make use of
\begin{equation}
\begin{split}
&\lim_{\lambda \rightarrow 0} \lambda^2 \int_0^{\lambda^{-2} t} \ud s_1 \int_0^{s_1} \ud s_2 \, \mathrm{e}^{\pm \mathrm{i} \omega_{1234} (s_1-s_2)} \\
&= t \int_0^\infty \ud s \, \mathrm{e}^{\pm \mathrm{i} \omega_{1234} s}
= t \, \left( \pm \mathrm{i} \, \mathcal{P}\Big(\frac{1}{\omega_{1234}} \Big) + \pi \, \delta(\omega_{1234}) \right)
\end{split}
\end{equation}
where $\mathcal{P}$ denotes the principal value. This yields
\begin{equation}
\begin{split}
\label{result}
&\lim_{\lambda \rightarrow 0} \lambda^2 \, W^{(2)}_\mathrm{c}(k,\lambda^{-2} \, t) \\
&= t \, \pi \int_{(\mathbb{T}^d)^3} \ud k_{234} \, \delta(\underline{k}) \, \delta(\omega_{1234})\\
&\hspace{40pt} \times
\langle \mathrm{f}, \, (\mathcal{A}[W]_{1234} + \mathcal{A}[W]^*_{1234}) \mathrm{g} \rangle \\ 
&\quad + \mathrm{i}\, t \int_{(\mathbb{T}^d)^3} \ud k_{234} \, \delta(\underline{k}) \, \mathcal{P}\Big(\frac{1}{\omega_{1234}} \Big) \\
&\hspace{40pt} \times \langle \mathrm{f}, \, (\mathcal{A}[W]_{1234} - \mathcal{A}[W]^*_{1234}) \mathrm{g} \rangle .
\end{split}
\end{equation}
%

We note that in case $W_{\sigma \tau}(k) = \delta_{\sigma \tau} W_\sigma(k)$ the term containing the principal part vanishes. 
The effective hamiltonian results from the $(2n+1)$-fold degeneracy of the unperturbed $H_0$.

\section{Simulation}
\label{sec:Simulation}

The details of the numerical implementation and mollification of the collision operator have been adapted from \cite{BoltzmannNonintegrable2013} to the bosonic case. Here we report simulation results. For better comparison we start always from the same initial state and modify the parameters of the evolution equation.

\subsection{Initial Wigner state}
\label{sec:InitialWigner}

\begin{figure*}[!ht]
\centering
\subfloat[matrix entries]{
\includegraphics[width=0.75\columnwidth]{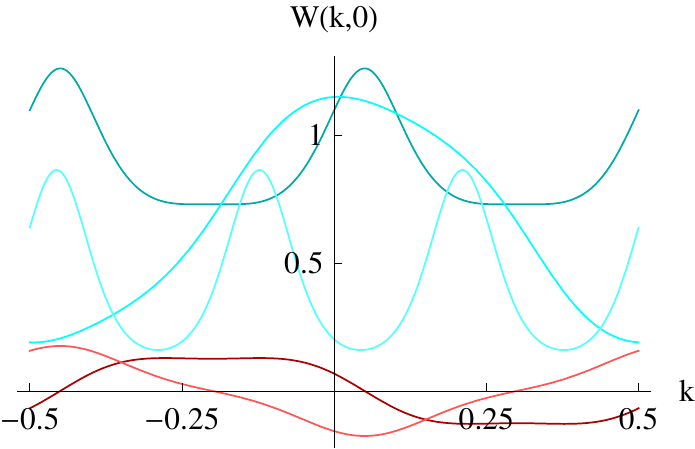}
\label{fig:W0matrix}} $\qquad$
\subfloat[eigenvalues]{
\includegraphics[width=0.75\columnwidth]{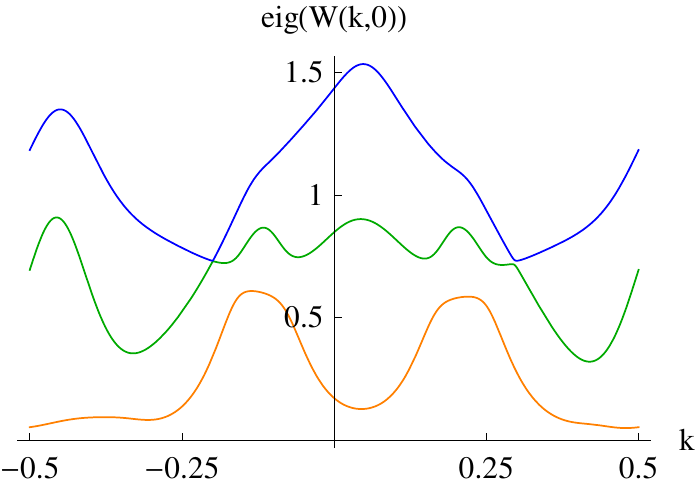}
\label{fig:W0eig}}
\caption{(Color online) The initial state $W(k,0)$ used for the simulations. (a) The cyan (upper) curves show the real diagonal entries, and the darker and lighter red curves the real and imaginary parts of the off-diagonal $\lvert0\rangle \langle\downarrow\rvert$ entry, respectively. For visual clarity, the remaining off-diagonal entries are omitted in the plot. (b) Eigenvalues of $W(k,0)$. Note the crossing (at negative $k$) and avoided crossing (at positive $k$) of the upper two curves.}
\label{fig:W0}
\end{figure*}

We fix the initial condition $W(k,0)$ as illustrated in Fig.~\ref{fig:W0}.
The cyan lines in Fig.~\ref{fig:W0matrix} represent the real diagonals, and the dark and light red functions the real and imaginary part of the off-diagonal $\ket{0}\bra{\downarrow}$ entry, respectively. The eigenvalues of $W(k,0)$ in Fig.~\ref{fig:W0eig} are non-negative for each $k \in \mathbb{T}$, as required, and $W(k,0)$ is continuous on $\mathbb{T}$. Note that the eigenvalues can exceed $1$, different from the Fermi case. It will be interesting to see how the eigenvalue crossing will evolve during the simulation.

\subsection{Stationary states}

\begin{figure*}[!ht]
\centering
\subfloat[nonthermal stationary state (nearest neighbor)]{
\includegraphics[width=0.8\columnwidth]{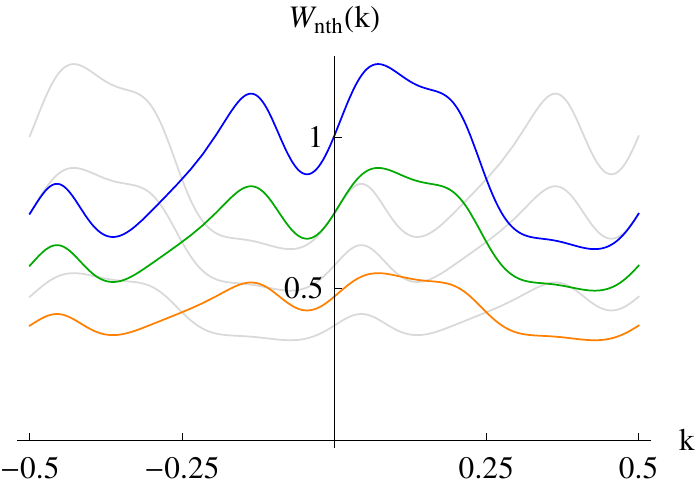}\label{fig:WstationaryNearest}} $\qquad$
\subfloat[corresponding $f$ function (nearest neighbor)]{
\includegraphics[width=0.8\columnwidth]{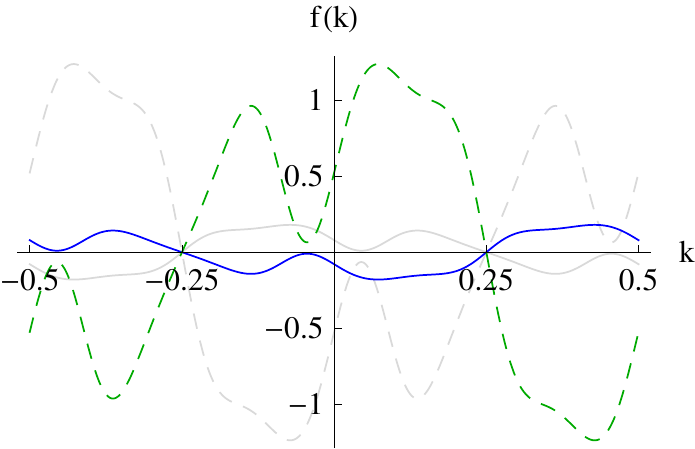}\label{fig:stationary_f}} \\
\subfloat[thermal equilibrium state (next-nearest neighbor hopping, with $\eta = \frac{1}{50}$)]{
\includegraphics[width=0.8\columnwidth]{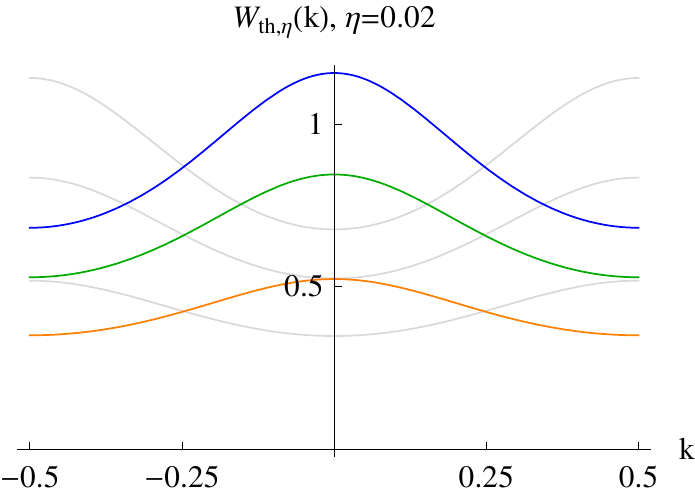}\label{fig:WstationaryNNN1}} $\qquad$
\subfloat[thermal equilibrium state (next-nearest neighbor hopping, with $\eta = \frac{1}{2}$)]{
\includegraphics[width=0.8\columnwidth]{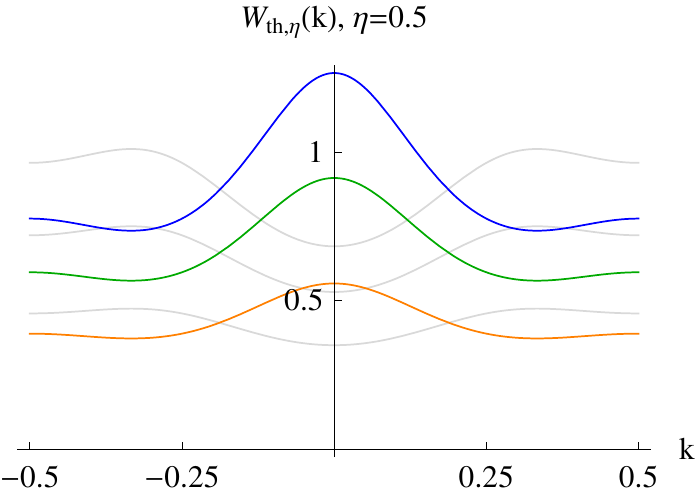}\label{fig:WstationaryNNN2}}
\caption{(Color online) Diagonal matrix entries (colored curves) of the stationary ($t \to \infty$) states corresponding to the initial $W(k,0)$ in Fig.~\ref{fig:W0}, for the nearest neighbor hopping model (a) and the next-nearest neighbor model with $\eta = \frac{1}{50}$ (c) and $\eta = \frac{1}{2}$ (d). The off-diagonal matrix entries are zero. Apparently, the final state sensitively depends on the dispersion relation $\omega(k)$. The additional conservation laws in the nearest neighbor case lead to a nonthermal stationary state. (b) The $f$ function (solid blue) and conserved $\mathrm{tr}[W(k)] - \mathrm{tr}[W(\tfrac12 - k)]$ (dashed green) for the nearest neighbor hopping model, which determines the nonthermal stationary state in (a). The faint gray curves show the corresponding entries when shifting the initial state $k \to k + \frac{1}{2}$, resulting in negative temperatures of the thermal stationary states. Note that the gray curves in (a) are exact shifted duplicates, which holds not true for the next-nearest neighbor models in (c) and (d).}
\label{fig:Wstationary}
\end{figure*}
One can obtain the stationary state corresponding to the initial $W(k,0)$ via the conservation laws Eq.~\eqref{eq:SpinConservation}, \eqref{eq:EnergyConservation} and \eqref{eq:TraceConservation}, as shown in Sec.~\ref{sec:StationarySolutions}. Different dispersion relations lead to different stationary states, which are illustrated in Fig.~\ref{fig:Wstationary} for the nearest and next-nearest neighbor hopping models. The next-nearest neighbor cases result in thermal Bose-Einstein distributions, while the nearest neighbor case results in a nonthermal stationary state of the form \eqref{eq:StationarySolutions}, see Fig.~\ref{fig:WstationaryNearest}. The corresponding $f$ function is shown in Fig.~\ref{fig:stationary_f}.

\subsection{Negative temperature}

States with negative temperatures ($\beta < 0$) have recently attracted interest \cite{Rosch2010, NegativeAbsoluteTemperature2013}. In our context, first observe that the exponential term of the Bose-Einstein distribution
\begin{equation}
\left( \mathrm{e}^{\beta(\omega(k) - \mu_\sigma)} - 1 \right)^{-1}
\end{equation}
is invariant under $\beta \to -\beta$ when simultaneously changing the sign of $\omega(k) - \mu_\sigma$. As argued in \cite{Rosch2010}, a sign flip of the \emph{nearest neighbor} dispersion (up to an arbitrary offset) is accomplished by shifting the momentum $k \to k + \tfrac{1}{2}$. In terms of the $f$ function in Eq.~\eqref{eq:StationarySolutions}, the shift of momentum is equivalent to a point reflection at the origin since $f(k + \frac{1}{2}) = -f(-k)$. However, for the next-nearest neighbor models the sign flip property of the dispersion holds not exactly true due to the additional $\eta \cos(4 \pi k)$ term, which is invariant under $k \to k + \frac{1}{2}$.

Nevertheless, it turns out that simply shifting the initial state in Fig.~\ref{fig:W0} by $k \to k + \frac{1}{2}$ suffices to obtain thermal equilibrium states with negative temperature. The states resulting from the initial shift are shown as faint gray curves in Fig.~\ref{fig:Wstationary}. Note that the thermal gray curves attain their maximum at (or close to) the boundary of the Brillouin zone, while positive temperature states have their maximum at $k = 0$. As expected, for the nearest neighbor model the $f$ function is reflected about the origin and the gray curves in (a) are shifted copies of the original colored curves, whereas for the next-nearest neighbor model this does no longer hold since $\omega_{\eta}(k + \frac{1}{2}) \neq -\omega_{\eta}(k) + c$ for nonzero $\eta$. The inverse temperature $\beta$ of the thermal states is shown in the following table. Note that the shift also changes the absolute value.
\begin{center}
\begin{tabular}{l|cc}
& $\eta = 0.02$ & $\eta = 0.5$ \\
\hline
$\beta$ of original $W(k,0)$ & \quad $0.1403$ & \ $0.1228$ \\
$\beta$ of shifted $W(k+\frac{1}{2},0)$ & $-0.1394$ & $-0.09507$ \\
\hline
\end{tabular}
\end{center}

\subsection{Time evolution and effect of the potential}

The three eigenvalues of a spin-$1$ Wigner state $W(k,t)$ define a point in $\mathbb{R}^3$.
\begin{figure}[!ht]
\centering
\includegraphics[width=0.8\columnwidth]{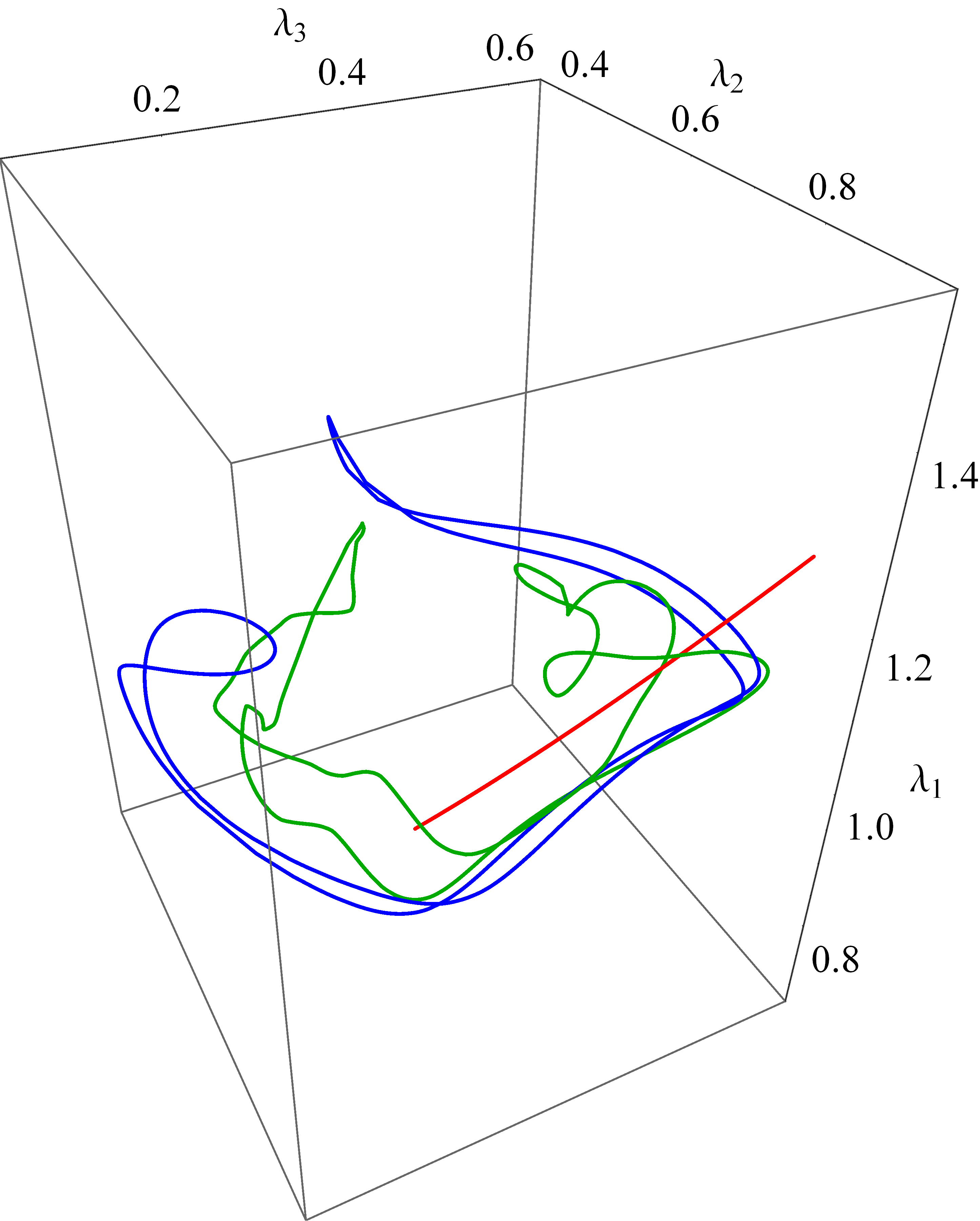}
\caption{Time evolution of the eigenvalues of $W(k,t)$ for the next-nearest neighbor model with $\eta = \frac{1}{2}$ and the $k$-dependent potential in Fig.~\ref{fig:Vpot}. Each curve shows the $3$ eigenvalues of $W(k,t)$ as $k$ traverses the Brillouin zone $\mathbb{T}$, for fixed $t$. The blue and green colors correspond to $t = 0$ (also see Fig.~\ref{fig:W0eig}) and $t = 1/16$, respectively. The red curve corresponds to the final thermal equilibrium state (illustrated in Fig.~\ref{fig:WstationaryNNN2}).}
\label{fig:WeigEvolvNNN2}
\end{figure}
We thus obtain for each $t$ a spectral curve of eigenvalues as $k$ traverses the Brillouin zone $\mathbb{T}$, as visualized in Fig.~\ref{fig:WeigEvolvNNN2} for the next-nearest neighbor model with $\eta = \frac{1}{2}$.

\begin{figure}[!ht]
\centering
\includegraphics[width=0.8\columnwidth]{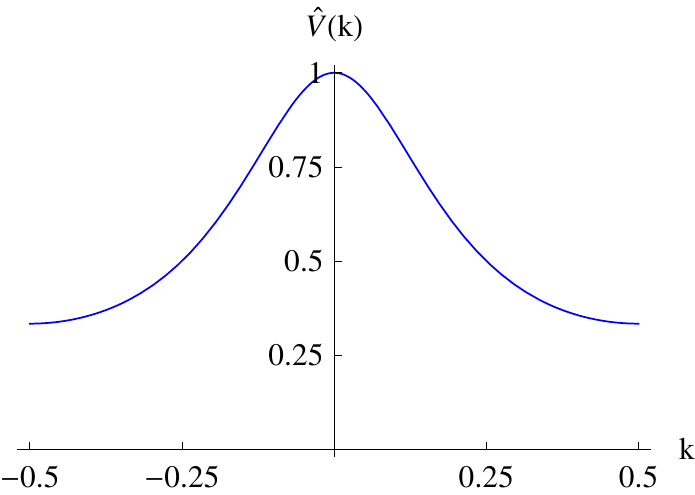}
\caption{The $k$-dependent potential $\hat{V}(k) = 1/(2 - \cos(2\pi k))$ used in the simulation in Fig.~\ref{fig:WeigEvolvNNN2}.}
\label{fig:Vpot}
\end{figure}
Comparing a simulation using the standard on-site potential $\hat{V}(k) \equiv 1$ with the $k$-dependent potential $\hat{V}(k) = 1/(2 - \cos(2\pi k))$, one notices that the convergence for the $k$-dependent potential is slower as compared to the on-site case; this observation can be confirmed quantitatively: the exponential decay rate in Hilbert-Schmidt norm is $0.67$ and $0.25$, respectively. The potential is visualized in Fig.~\ref{fig:Vpot}.

\begin{figure*}[!ht]
\centering
\subfloat[$\mathrm{eig}(\mathcal{A}_{\mathrm{quad}})$]{
\includegraphics[width=0.8\columnwidth]{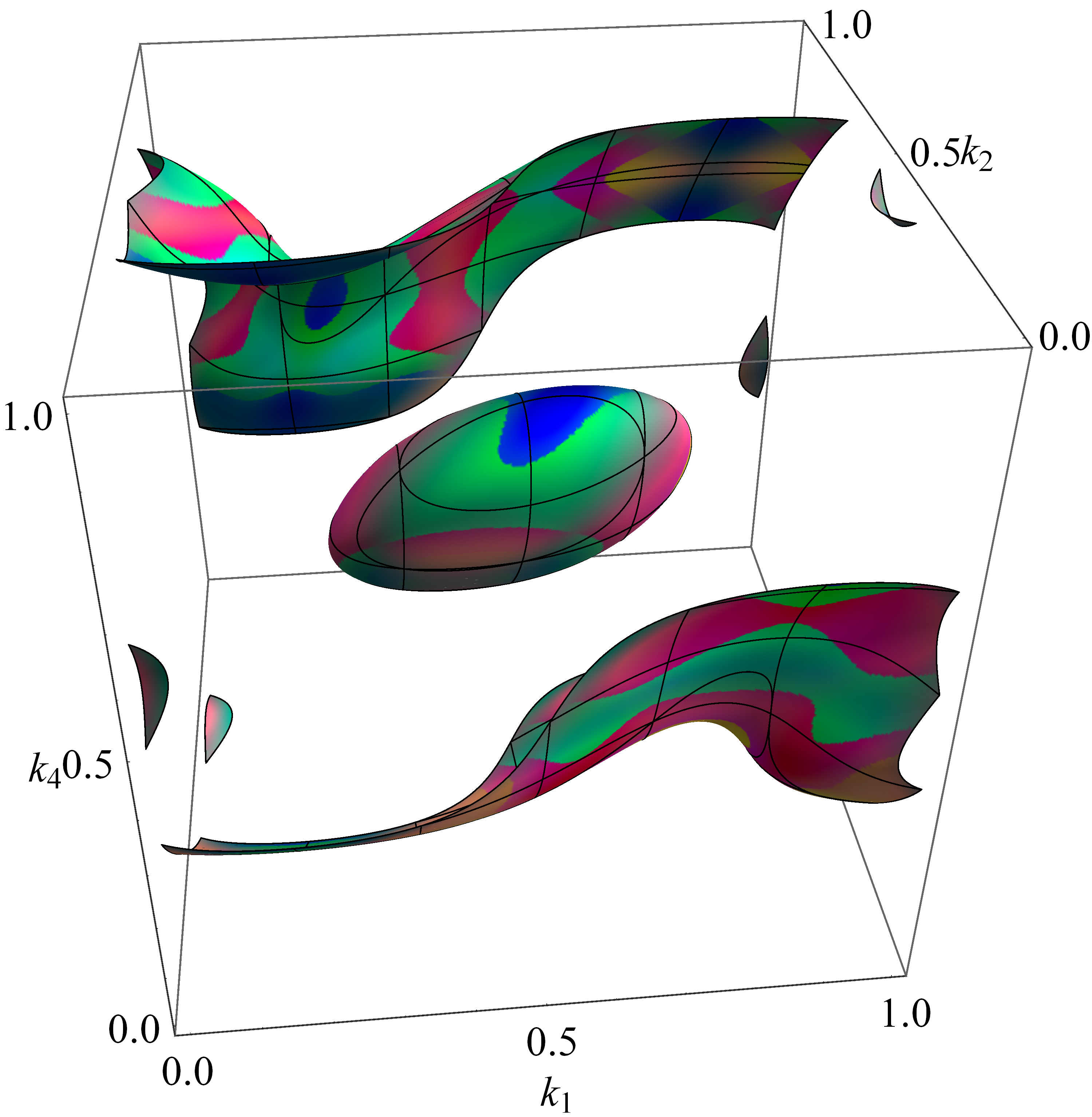}}
\qquad
\subfloat[$\mathrm{eig}(\mathcal{A}_{\mathrm{tr}})$]{
\includegraphics[width=0.8\columnwidth]{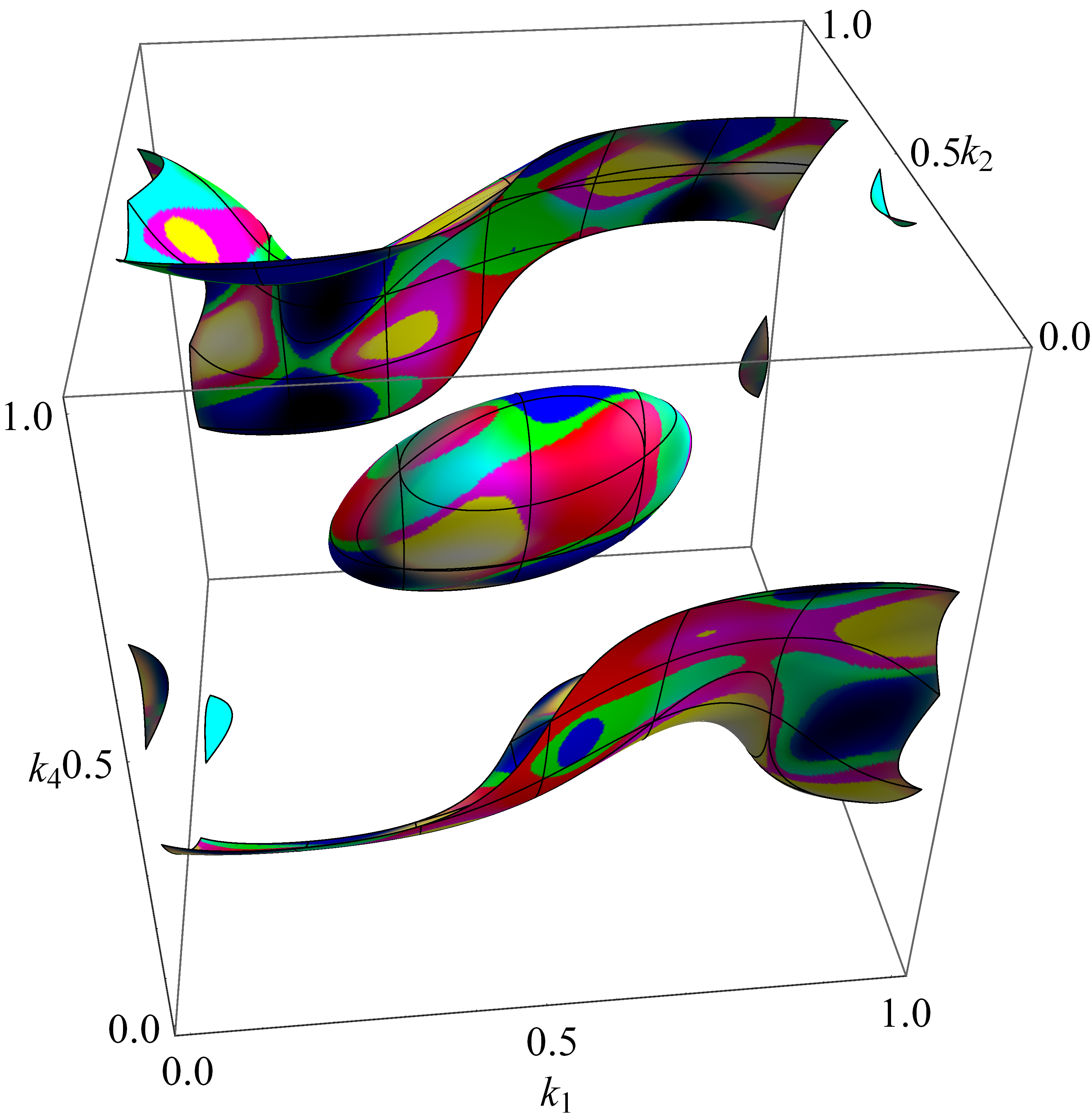}}
\caption{(Color online) Three-dimensional shape of the $\gamma_{\mathrm{diag}}$ and
$\gamma_{\mathrm{ellip}}$ collision manifolds for the next-nearest neighbor model with $\eta = \frac{1}{2}$. Color encodes the eigenvalues of (a) $\mathcal{A}_{\mathrm{quad}}$ and (b) $\mathcal{A}_{\mathrm{tr}}$ in Eq.~\eqref{eq:Aquad} and \eqref{eq:Atr} with the $\hat{V}_{ij}$ prefactors set to $1$, for the initial state $W(k,0)$. Eigenvalues can be negative, and the zero state corresponds to gray color.}
\label{fig:collision_manifold}
\end{figure*}

The kinematically allowed collisions $\delta(\underline{k})\,\delta(\underline{\omega})$ define the collision manifold, a subset of $\mathbb{T}^4$. Specifically for the next-nearest neighbor model with $\eta = \frac{1}{2}$, it consists of the $\gamma_1$, $\gamma_2$, $\gamma_{\mathrm{diag}}$ and $\gamma_{\mathrm{ellip}}$ manifolds as discussed in \cite{BoltzmannNonintegrable2013}. Fig.~\ref{fig:collision_manifold}
shows the latter two, with color encoding the eigenvalues of $\mathcal{A}_{\mathrm{quad}}$ on the left and $\mathcal{A}_{\mathrm{tr}}$ on the right (for the initial state $W(k,0)$ and $\hat{V}(k) \equiv 1$). Considering the effect of the potential in Fig.~\ref{fig:Vpot}, let us briefly elaborate on the weighting of the collisions by the $\hat{V}$-prefactors of the $\mathcal{A}_{\mathrm{quad}}$ and $\mathcal{A}_{\mathrm{tr}}$ integrands. Since $\hat{V}(k)$ attains its maximum at $k = 0$, the scale factor $\hat{V}(k_2 - k_3) \hat{V}(k_3 - k_4)$ is largest when the momenta $k_1, \dots, k_4$ are all equal. Concerning $\hat{V}(k_3 - k_4)^2$, the hyperplane $k_3 = k_4$ (or equivalently $k_1 = k_2$) contributes the most.

\subsection{Exponential convergence and prethermalization}

\begin{figure*}[!ht]
\centering
\subfloat[entropy increase]{
\includegraphics[width=0.9\columnwidth]{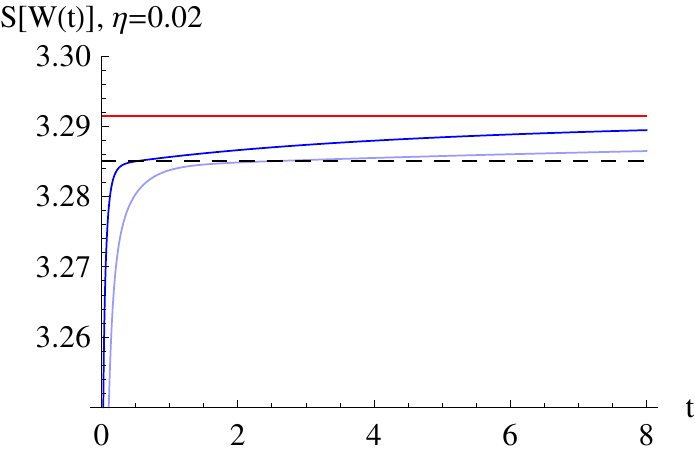}
\label{fig:entropy_conv_eta1}
} $\qquad$
\subfloat[convergence of the off-diagonal entries]{
\includegraphics[width=0.9\columnwidth]{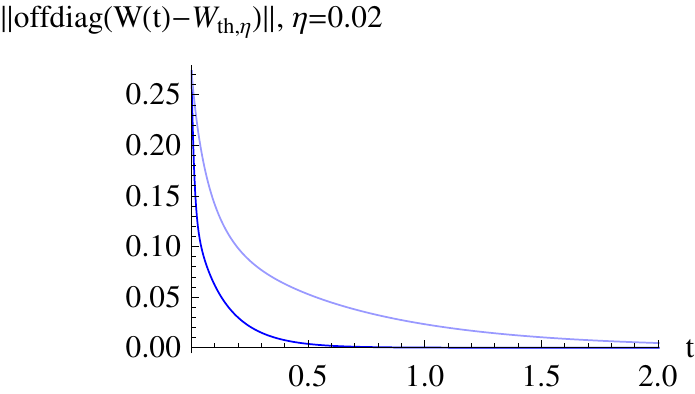}
\label{fig:offdiag_conv_eta1}}
\caption{(Color online) (a) Entropy increase for the next-nearest neighbor model with small $\eta = \frac{1}{50}$ (dark blue curve). The red curve shows the entropy of the corresponding equilibrium state, and the dashed black curve the entropy of the stationary nearest neighbor state. The entropy increases quickly up to $t \simeq 0.5$, where it reaches the dashed curve (``fast motion''). Afterwards it slowly approaches the actual thermal equilibrium value (``slow motion''). (b) Exponential convergence of the off-diagonal entries. The dynamic matches the ``fast motion'' in (a) quite well, i.e., the off-diagonal entries (almost) reach zero within the ``fast motion'' period. For visual clarity, the time axis in (b) is shorter than in (a). To demonstrate the effect of the potential, the faint blue curves show the results for a calculation with the potential in Fig.~\ref{fig:Vpot} instead of the uniform $\hat{V}(k) \equiv 1$.}
\label{fig:convergence_eta1}
\end{figure*}
The next-nearest neighbor model with small $\eta = \frac{1}{50}$ serves as illustration of the prethermalization effect. In our context, the initial Wigner state converges quickly to a quasistationary state close to the nonthermal stationary state in Fig.~\ref{fig:WstationaryNearest} (nearest neighbor model with $\eta = 0$), and then thermalizes slowly to the equilibrium state in Fig.~\ref{fig:WstationaryNNN1}. The entropy increase (shown in Fig.~\ref{fig:entropy_conv_eta1}) quantifies this dynamical picture: the entropy quickly reaches the entropy of the stationary nearest neighbor state (dashed black curve), and then further increases towards the actual thermal equilibrium state. An analytical approach in terms of the vanishing off-diagonal entries follows the same lines as in Ref.~\cite{BoltzmannNonintegrable2013}, and is illustrated in Fig.~\ref{fig:offdiag_conv_eta1}.

\section{Conclusions}
\label{sec:Conclusions}

On the kinetic level, the dynamics of bosons and fermions in one dimension is qualitatively similar: additional conservation laws and nonthermal stationary states exist for pure nearest neighbor hopping. These additional conservation laws disappear when turning on longer range hopping terms, and all stationary states become thermal equilibrium states. Prethermalization is observed for small next-nearest neighbor hopping.

Conversely, the main modifications for bosons include the following: $\tilde{W} = 1 - W$ for fermions is replaced by $\tilde{W} = 1 + W$ for bosons, the Wigner matrix $W(k)$ has dimension $(2n+1) \times (2n+1)$ where $n \in \mathbb{N}_0$ is the spin quantum number, and the Fermi property $0 \leq W(k) \leq 1$ is relaxed to $0 \leq W(k)$ for bosons.

Concerning negative temperatures, we have demonstrated that a simple shift $k \to k + \frac{1}{2}$ in the initial state suffices to change the temperature sign of the corresponding thermal equilibrium state. In this context, the shift-invariance of the evolution dynamics with respect to $k \to k + \frac{1}{2}$ is broken by the dispersion relation whenever $\omega(k + \frac{1}{2}) \neq -\omega(k) + c$.

On the microscopic level the Fermi-Hubbard hamiltonian with on-site potential and nearest neighbor coupling is integrable and has an infinite number of conservation laws. In \cite{BoltzmannNonintegrable2013} we concluded that this integrable structure is still visible on the kinetic level. The spin-$0$ Bose-Hubbard hamiltonian, with the same couplings, is not integrable, but the Boltzmann transport equation still has an infinite number of conservation laws. In the $t$-$V$ limit integrability of the Hubbard hamiltonian is regained at the expense of the occupation numbers taking values $0, 1$ only.  This constraint is not readily transcribed to a transport equation. We infer that the link between microscopic integrability and infinite number of conservation laws on the kinetic level is less stringent than anticipated in \cite{BoltzmannNonintegrable2013}.

\appendix

\section{Positivity}
\label{sec:Positivity}

The following lemma ensures positivity of the gain term in Eq.~\eqref{eq:AW1234def}, when identifying $x = \hat{V}_{34}$, $y = - \hat{V}_{23}$ and using the interchangeability of the integration variables $k_2 \leftrightarrow k_4$.
\begin{lemma}
Let $A, B, C \in \mathbb{C}^{d \times d}$ be positive semidefinite and $x, y \in \mathbb{R}$. Then
\begin{equation*}
x^2 \, A\, \mathrm{tr}[B\,C] + y^2\, C\, \mathrm{tr}[B\,A] - x\,y \, A B C - x\,y \, C B A \ge 0.
\end{equation*}
\begin{proof}
By the spectral decomposition of $B$ with non-negative eigenvalues, we can without loss of generality assume that $B = \ket{\psi}\bra{\psi}$ for a $\psi \in \mathbb{C}^d$. Now let $\varphi \in \mathbb{C}^d$ be arbitrary, then
\begin{equation*}
\begin{split}
&\bra{\varphi} x^2 \, A\, \mathrm{tr}[B\,C] + y^2\, C\, \mathrm{tr}[B\,A] - x\,y \, A B C - x\,y \, C B A \ket{\varphi}\\
&= x^2 \bra{\varphi} A \ket{\varphi} \bra{\psi} C \ket{\psi} + y^2 \bra{\varphi} C \ket{\varphi} \bra{\psi} A \ket{\psi}\\
&\quad - x\,y \bra{\varphi} A \ket{\psi} \bra{\psi} C \ket{\varphi} - x\,y \bra{\varphi} C \ket{\psi} \bra{\psi} A \ket{\varphi}\\
&\ge x^2 \bra{\varphi} A \ket{\varphi} \bra{\psi} C \ket{\psi} + y^2 \bra{\varphi} C \ket{\varphi} \bra{\psi} A \ket{\psi}\\
&\quad - 2\,\abs{x\,y} \cdot \abs{\bra{\varphi} A \ket{\psi}} \cdot \abs{\bra{\psi} C \ket{\varphi}}.
\end{split}
\end{equation*}
Using the Cauchy-Schwarz inequality $\abs{\bra{\varphi} A \ket{\psi}}^2 \le \bra{\varphi} A \ket{\varphi} \bra{\psi} A \ket{\psi}$, we arrive at the further estimate
\begin{equation*}
\begin{split}
&\ge x^2 \bra{\varphi} A \ket{\varphi} \bra{\psi} C \ket{\psi} + y^2 \bra{\varphi} C \ket{\varphi} \bra{\psi} A \ket{\psi}\\
&\quad - 2\,\abs{x\,y} \sqrt{\bra{\varphi} A \ket{\varphi} \bra{\psi} A \ket{\psi}} \sqrt{\bra{\varphi} C \ket{\varphi} \bra{\psi} C \ket{\psi}}\\
& = \left(\abs{x} \sqrt{\bra{\varphi} A \ket{\varphi} \bra{\psi} C \ket{\psi}} - \abs{y} \sqrt{\bra{\varphi} C \ket{\varphi} \bra{\psi} A \ket{\psi}}\right)^2\\
&\ge 0.
\end{split}
\end{equation*}

\end{proof}

\end{lemma}

\section{Bosonic correlations}
\label{sec:BosonicCorrelations}

\subsection{Two-point function}

Let $\mathsf{H} = \sum_{k,l \in \mathbb{Z}} H_{kl} \, a_k^* a_l$ be the second quantization of the one-particle matrix $H$. It is assumed that $\mathrm{e}^{-H}$ is trace class and $\det(1 + \mathrm{e}^H) \neq 0$. We use the identities
\begin{equation}
\mathrm{e}^{-\mathsf{H}} a_i^* \mathrm{e}^{\mathsf{H}} = \sum_{j \in \mathbb{Z}} a_j^* \big(\mathrm{e}^{-H}\big)_{ji}, \quad \mathrm{e}^{-\mathsf{H}} a_i \mathrm{e}^{\mathsf{H}} = \sum_{j \in \mathbb{Z}} \big(\mathrm{e}^{H}\big)_{ij} a_j. 
\end{equation}
Then
\begin{equation}
\begin{split}
\langle a_i^* a_j \rangle 
&= \frac{1}{Z} \mathrm{tr}\big[\mathrm{e}^{- \mathsf{H}} a_i^* a_j\big] 
= \sum_n \frac{1}{Z} \mathrm{tr}\big[a_n^* \big(\mathrm{e}^{-H}\big)_{ni} \mathrm{e}^{-\mathsf{H}} a_j\big] \\
&= \sum_n \frac{1}{Z} \mathrm{tr}\big[ \big(\mathrm{e}^{-H}\big)_{ni} \mathrm{e}^{-\mathsf{H}} a_j a_n^*\big] \\
&= \sum_n \big(\mathrm{e}^{-H}\big)_{ni} \frac{1}{Z} \mathrm{tr}\big[\mathrm{e}^{-\mathsf{H}} a_j a_n^*\big] \\
&= \sum_n \big(\mathrm{e}^{-H}\big)_{ni} \frac{1}{Z} \mathrm{tr}\big[\mathrm{e}^{-\mathsf{H}} (\delta_{nj} + a_n^* a_j)\big] \\
&= \big(\mathrm{e}^{-H}\big)_{ji} + \sum_n \langle a_n^* a_j \rangle \big(\mathrm{e}^{-H}\big)_{ni}
\end{split}
\end{equation}
with the partition function $Z = \mathrm{tr}[\mathrm{e}^{- \mathsf{H}}]$. Rearranging gives
\begin{equation}
\sum_{n \in \mathbb{Z}} \big\langle a_n^* \big(1 - \mathrm{e}^{- H}\big)_{ni} a_j \big\rangle = \big(\mathrm{e}^{-H}\big)_{ji}.	
\end{equation}
Finally multiplying this expression by $\left( (1 - \mathrm{e}^{-H})^{-1} \right)_{im}$ and summing over the $i$ variable, we obtain
\begin{equation}
\langle a_m^* a_j \rangle = \big((\mathrm{e}^{H}-1)^{-1}\big)_{jm}.	
\end{equation}

\subsection{Expansion as permanent}

We prove recursively that
\begin{equation} \label{eq:perm1}
\langle a_{i_1}^* a_{j_1} \cdots a_{i_n}^* a_{j_n} \rangle = \mathrm{perm}[K(i_k,j_l)]_{1 \leq k, l \leq n},	
\end{equation}
where
\begin{equation}
K(i_k,j_l) = \begin{cases}
	\langle a_{i_k}^* a_{j_l} \rangle & \text{if } k \leq l, \\
	\langle a_{l_l} a_{i_k}^* \rangle & \text{if } k > l. \end{cases}
\end{equation}
%
For $n = 1$ the formula bolds by definition. Suppose the formula \eqref{eq:perm1} has been established for some $n$, i.e.,
\begin{multline}
\label{eq:perm2}
\langle a_{i_1}^* a_{j_1} \cdots a_{i_n}^* a_{j_n} \rangle \\
= \mathrm{perm}\! \begin{bmatrix}
	\langle a_{i_1}^* a_{j_1} \rangle & \langle a_{i_1}^* a_{j_2} \rangle & \cdots & \langle a_{i_1}^* a_{j_n} \rangle \\
	\langle a_{j_1} a_{i_2}^* \rangle & \langle a_{i_2}^* a_{j_2} \rangle & \cdots & \langle a_{i_2}^* a_{j_n} \rangle \\
	\vdots & \vdots & \ddots & \vdots \\
	\langle a_{j_1} a_{i_n}^* \rangle & \langle a_{j_2} a_{i_n}^* \rangle & \cdots & \langle a_{i_n}^* a_{j_n} \rangle
\end{bmatrix}.
\end{multline}
We will need one more expression for $\langle \cdots \rangle$ such that in the first $k$ pairs the annihilation operator precedes the creation operator,
\begin{equation} \label{eq:det3}
\begin{split}
&\langle a_{j_1} a_{i_1}^* \cdots a_{j_k} a_{i_k}^* a_{i_{k+1}}^* a_{j_{k+1}} \cdots a_{i_n}^* a_{j_n} \rangle = \mathrm{perm} \\
& \begin{bmatrix}
\langle a_{j_1}^* a_{i_1} \rangle & \cdots & \langle a_{i_1}^* a_{j_k} \rangle & 
			\langle a_{i_1}^* a_{j_{k+1}} \rangle & \cdots & \langle a_{i_1}^* a_{j_n} \rangle \\
\vdots & \ddots & \vdots & \vdots & \ddots & \vdots \\	
\langle a_{j_k} a_{i_1}^* \rangle & \cdots & \langle a_{j_k} a_{i_k}^* \rangle & 
			\langle a_{i_k}^* a_{j_{k+1}} \rangle & \cdots & \langle a_{i_k}^* a_{j_n} \rangle \\
\langle a_{j_{k+1}} a_{i_1}^* \rangle & \cdots & \langle a_{j_{k+1}} a_{i_k}^* \rangle & 
			\langle a_{i_{k+1}}^* a_{j_{k+1}} \rangle & \cdots & \langle a_{i_{k+1}}^* a_{j_n} \rangle \\
\vdots & \ddots & \vdots & \vdots & \ddots & \vdots \\
\langle a_{j_n} a_{i_1}^* \rangle & \cdots & \langle a_{j_n} a_{i_k}^* \rangle & 
			\langle a_{j_n} a_{i_{k+1}}^* \rangle & \cdots & \langle a_{i_n}^* a_{j_n} \rangle
\end{bmatrix}.
\end{split}
\end{equation}
Let us proof this formula. For $k = 0$, it agrees with \eqref{eq:perm2}. Suppose it to be true for some $k$. Let us then prove that the formula \eqref{eq:det3} holds for $k+1$,
\begin{equation}
\label{eq:perm4}
\begin{split}
&\langle a_{j_1} a_{i_1}^* \cdots a_{j_{k+1}} a_{i_{k+1}}^* a_{i_{k+2}}^* a_{j_{k+2}} \cdots a_{i_n}^* a_{j_n} \rangle \\
& = \langle a_{j_1} a_{i_1}^* \cdots a_{j_k} a_{i_k}^* a_{i_{k+1}}^* a_{j_{k+1}} \cdots a_{i_n}^* a_{j_n} \rangle \\
&\quad + \delta_{i_{k+1},j_{k+1}} \langle a_{j_1} a_{i_1}^* \cdots a_{j_k} a_{i_k}^* a_{i_{k+2}}^* a_{j_{k+2}} \cdots a_{i_n}^* a_{j_n} \rangle. 
\end{split}
\end{equation}
Using the expression \eqref{eq:det3} and considering the expansion of the permanent in the $(k+1)^{\mathrm{th}}$ column (or row), it is easy to see that \eqref{eq:perm4} corresponds to the expression \eqref{eq:det3} but with the diagonal term $a_{i_{k+1}}^* a_{j_{k+1}}$ replaced by $a_{j_{k+1}} a_{i_{k+1}}^*$. Therefore \eqref{eq:det3} holds for $k+1$, too.

Now we prove \eqref{eq:perm2} for $n+1$ by using \eqref{eq:perm2} for $n$ and \eqref{eq:det3} for $n$ and $k \leq n$,
\begin{equation*}
\begin{split}
&\langle a_q^* a_{j_1} \cdots a_{i_{n+1}}^* a_{j_{n+1}} \rangle \\
&= \frac{1}{Z} \mathrm{tr}\big[ \mathrm{e}^{-\mathsf{H}} a_q^* a_{j_1} \cdots a_{i_{n+1}}^* a_{j_{n+1}}\big] \\
&= \sum_{n \in \mathbb{Z}} \frac{1}{Z} \big(\mathrm{e}^{-H}\big)_{mq} \mathrm{tr}\big[\mathrm{e}^{-\mathsf{H}} a_{j_1} \cdots a_{i_{n+1}}^* a_{j_{n+1}} a_m^*\big] \\
&= \sum_{m \in \mathbb{Z}} \big(\mathrm{e}^{-H}\big)_{mq} \langle a_m^* a_{j_1} \cdots a_{i_{n+1}}^* a_{j_{n+1}} \rangle \\
&\quad + \sum_{p = 2}^{n+1} \big(\mathrm{e}^{-H}\big)_{j_p q} \\
&\qquad \times \langle a_{j_1} a_{i_2}^* \cdots a_{j_{p-1}} a_{i_p}^* a_{i_{p+1}}^* a_{j_{p+1}} \cdots a_{i_{n+1}}^* a_{j_{n+1}} \rangle \\
&\quad + \big(\mathrm{e}^{-H}\big)_{j_1 q} \langle a_{i_2}^* a_{j_2} \cdots a_{i_{n+1}}^* a_{j_{n+1}} \rangle.
\end{split}
\end{equation*}
We take the term with the sum over $m \in \mathbb{Z}$ together with the first one and multiply the whole expression by $\sum_{q \in \mathbb{Z}} \big((1 - \mathrm{e}^{-H})^{-1}\big)_{q i_1}$ to obtain
\begin{multline}
\langle a_{i_1}^* a_{j_1} \cdots a_{i_{n+1}}^* a_{j_{n+1}} \rangle 
	= \langle a_{i_1}^* a_{j_1} \rangle \langle a_{i_2}^* a_{j_2} \cdots a_{i_{n+1}}^* a_{j_{n+1}} \rangle \\
	+ \sum_{p = 2}^{n+1} \langle a_{i_1}^* a_{j_p} \rangle 
		\langle a_{j_1} a_{i_2}^* \cdots a_{j_{p-1}} a_{i_p}^* a_{i_{p+1}}^* a_{j_{p+1}} \cdots a_{i_{n+1}}^* a_{j_{n+1}} \rangle.
\end{multline}
Using \eqref{eq:perm2} and \eqref{eq:det3} for $n$ terms we see that this last expression is nothing else than the expansion with respect to the first row of \eqref{eq:perm2} with $n$ substituted by $n+1$.


\end{document}